\documentclass[reqno,12pt]{article}
\usepackage{hyperref,cite}
\usepackage{hyperref,amsfonts,amssymb,amsmath,amscd,bm,bbold,
delarray,subfigure}
\usepackage{xypic}
\usepackage[dvips]{color}
\usepackage[dvips]{graphicx}

\DeclareMathOperator{\ad}{ad}

\DeclareMathOperator{\im}{im}
\DeclareMathOperator{\id}{id}
\DeclareMathOperator{\Hom}{Hom}

\DeclareMathOperator{\pro}{pr}

\DeclareMathOperator{\Vrtc}{Vert}
\DeclareMathOperator{\End}{End}

\DeclareMathOperator{\Lie}{Lie}
\DeclareMathOperator{\ann}{ann}
\DeclareMathOperator{\rank}{rank}

\numberwithin{equation}{section}

\font\sansserif=cmss12
\font\scriptsansserif=cmss12 at 7 truept
\font\scriptscriptsansserif=cmss10 at 5 truept
\font\euler=eusm10 at 12 truept
\font\scripteuler=eusm7
\font\scriptscripteuler=eusm5 
\def\eul{\fam=12}
\newcommand{\matheul}[1]{{{\eul #1}}}

\newcommand{\mathbfs}[1]{{\boldsymbol{#1}}}

\newcommand{\Beta}{\mathrm{B}}

\begin{document}

\hrule\vskip.4cm
\hbox to 14.5 truecm{October 2008\hfil DFUB 08}
\hbox to 14.5 truecm{Version 1  \hfil } 
\vskip.4cm
\hrule
\vskip.7cm
\begin{large}
\centerline{\textcolor{blue}{\bf The Lie algebroid Poisson sigma model} }  
\end{large}
\vskip.2cm
\centerline{by}
\vskip.2cm
\centerline{\textcolor{blue}{\bf\bf Roberto Zucchini}}
\centerline{\it Dipartimento di Fisica, Universit\`a degli Studi di Bologna}
\centerline{\it V. Irnerio 46, I-40126 Bologna, Italy}
\centerline{\it I.N.F.N., sezione di Bologna, Italy}
\centerline{\it E--mail: zucchinir@bo.infn.it}
\vskip.7cm
\hrule
\vskip.7cm
\centerline{\textcolor{blue}{\bf Abstract}}
\par\noindent
\vskip.2cm
The Poisson--Weil sigma model, worked out by us in \cite{Zucchini6,Zucchini7}, 
stems from gauging a Hamiltonian Lie group symmetry of the target 
space of the Poisson sigma model. 
Upon gauge fixing of the BV master action, it yields interesting topological field theories such as
the 2--dimensional Donaldson-Witten topological gauge theory and the gauged A 
topological sigma model. 
In this paper, generalizing the above construction, we construct the Lie algebroid 
Poisson sigma model. This is yielded by gauging a Hamiltonian Lie groupoid symmetry 
of the Poisson sigma model target space. 
We use the BV quantization approach in the AKSZ geometrical version to ensure 
consistent quantization and target space covariance. 
The model has an extremely rich geometry and an intricate BV cohomology,
which are studied in detail.


\vfill\eject
\tableofcontents
\vfill\eject

\section{\normalsize \textcolor{blue}{Introduction}}\label{sec:intro}

In geometry and in physics, symmetry is normally described in terms of groups and group
actions. However, there are more general forms of symmetry, which do not let themselves 
be dealt with in that way, but which, nevertheless, are clearly recognizable as such.
The mathematical structure that underlies them is that of groupoid and groupoid action.

The algebraic notion of groupoid was introduced by W. Brandt in 1926 as a generalization of 
that of group \cite{Brandt1}. Since then, groupoids have found a wide range of mathematical 
applications. Topological and Lie groupoids, groupoids equipped with a compatible topological and 
differential structure, were used systematically by Ehresmann in his work in algebraic and 
differential geometry \cite{Ehresmann1}. Groupoids were also employed in algebraic geometry by Grothendieck
\cite{Grothendieck1} and in algebraic topology by Brown 
\cite{Brown1}. 
The notion of principal bundles with structure groupoid was worked out by Connes 
in the study of the holonomy groupoid of a foliation in \cite{Connes1}.
Coste, Dazord, Weinstein, Karasev and Zakrzewski
used symplectic groupoids, Lie groupoids equipped with a compatible symplectic structure,
in the study of non commutative deformations of the algebra of smooth functions on a 
manifold \cite{Weinstein2,Weinstein3,Karasev1,Zakrzewski1}.
In \cite{Weinstein4}, Weinstein introduced Poisson groupoids
as generalizations of both Poisson Lie groups and symplectic groupoids.
See \cite{Weinstein1} for a review of applications of groupoids in mathematics.

Lie algebroids were first studied by Pradines in the early sixties in relation with Lie 
groupoids. \cite{Pradines1}. Since then, they have proved to provide a very general and 
flexible framework for studying a wide range of geometrical structures. 

Lie algebroids are a vector bundle generalization of Lie algebras. A Lie algebra is just a Lie 
algebroid over a point. To any Lie groupoid there is associated a Lie algebroid much in the same way 
as to a Lie group there is associated a Lie algebra. Therefore, Lie algebroid theory parallels
Lie groupoid theory as the infinitesimal version of the latter.
However, the scope of 
Lie algebroids is broader, since, unlike what happens for ordinary Lie algebras,
not every Lie algebroid is integrable to a Lie groupoid. The conditions for integrability were found 
in \cite{Crainic1}.

In recent years, Lie groupoids and Lie algebroids have attracted much interest also in theoretical physics 
because of their potential of describing the generalized forms of symmetry arising in the so called non 
linear gauge theories \cite{Schoutens1}.
In ordinary linear gauge theories, the symmetries are local, the symmetry algebra closes off-shell and the 
symmetry algebra structure constants are field independent. By contrast, 
in non linear gauge theories, the symmetries are still local, but the symmetry algebra closes only on-shell
and the symmetry algebra structure constants are field dependent (and, so, actually structure functions). 
While the symmetries of linear gauge theories are amenable by standard Lie theoretic techniques
in an essentially finite dimensional setting, those of non linear gauge theories are not manifestly so. 
This renders the geometrical properties of non linear gauge theories rather mysterious 
and poses serious 
problems for their consistent quantization. 

Lie groupoids and Lie algebroids constitute a promising framework for studying the 
symmetries of non linear gauge theories \cite{Levin1,Strobl3}. 
They can accommodate the local symmetries of these theories 
while still allowing for an essentially finite dimensional treatment. Moreover, they reproduce
the standard Lie theoretic framework in the linear case. 

This approach, though completely general, has been adopted mostly in the study of the Poisson 
sigma model \cite{Ikeda2,Strobl1}, the prototype non linear gauge theory, and related models. 
In \cite{ Strobl2,Strobl4,Strobl5}, the field equations of these field theories are interpreted 
as morphisms from the space--time tangent Lie algebroid to a certain target Lie algebroid and their 
on-shell symmetries as homotopies of such morphisms. In \cite{Cattaneo4,Bonechi3}, a Poisson sigma model 
with an integrable Poisson target manifold is considered, the field equations are interpreted 
as morphisms from the world sheet fundamental groupoid to the target manifold integrating groupoid
and the symmetries are described in terms of an infinite dimensional 
groupoid of maps from the world sheet to this groupoid. In \cite{Stienon1}, these constructions have 
been interpreted in terms of the geometry of appropriate principal groupoid bundles.
The present work is one more step in the same direction, but from a different perspective. 

The Poisson--Weil sigma model, introduced by us in ref.
\cite{Zucchini6} and further refined in \cite{Zucchini7} 
(see also \cite{Signori1}), 
is a Poisson sigma model in which a Hamiltonian Lie group symmetry of the target space is gauged. 
In \cite{Zucchini7}, it is shown that, upon carrying out an appropriate gauge fixing, 
the Poisson--Weil model yields the 2--dimensional version of Donaldson–-Witten topological gauge theory, 
describing the moduli space of flat connections on a closed surface \cite{Witten5,Witten6},
in the pure gauge case where the target space is a point, and the gauged A topological 
sigma model describing the moduli space of solutions of 
the so called vortex equations  worked out by Baptista 
\cite{Baptista1,Baptista2,Baptista3}, in the case where the target space 
is a manifold with a Kaehler structure preserved by the symmetry action.
In this paper, developing on the results of \cite{Zucchini6,Zucchini7}, we construct the 
Lie algebroid Poisson sigma model. In simple terms, this is a Poisson sigma model in which a 
Hamiltonian Lie groupoid symmetry of the target space is gauged.
In more precise terms, we consider the Poisson sigma model on a Poisson manifold $X$ fibered over 
another manifold $M$ and gauge the symmetry associated with a Hamiltonian infinitesimal 
action of a regular Lie algebroid $L$ over $M$ on $X$.
Though we have in mind especially the case where $L$ is the Lie algebroid $AG$ of a Lie groupoid $G$,
the model is consistently defined also when $L$ is not integrable.
The whole construction is conceived in such a way to reproduce the Poisson--Weil 
sigma model of refs. \cite{Zucchini6,Zucchini7} 
in the particular case where $M$ is a point and $L$ is an ordinary Lie algebra.

The model has a rich geometry. The kernel $\ker\rho$ of the anchor $\rho$ of
$L$ plays a central role in the whole construction. The target space geometry involves a generalized 
moment map, which is defined only on sections of $\ker\rho$. When the Lie algebroid $L$ integrates 
to a Lie groupoid $G$, the model encodes a generalized fiberwise Hamiltonian reduction
of the target Poisson manifold $X$ \cite{Marsden1,Ratiu1}. 

We use the Batalin--Vilkoviski (BV) quantization approach \cite{BV1,BV2}
in the geometrical version of Alexandrov, Kontsevich, Schwartz and Zaboronsky (AKSZ) 
\cite{AKSZ}, as developed by Cattaneo and Felder in \cite{Cattaneo1,Cattaneo2}.
The BV cohomology turns out to be quite intricate, intertwining as it does the Lie algebroid cohomology of $L$
and the tangential Poisson cohomology of $X$.

The hope of our work is to find sensible gauge fixing prescriptions of the Lie algebroid Poisson sigma model
yielding interesting topological field theories generalizing the ones mentioned above
(see \cite{Bonechi3} for an attempt in this direction). We leave this for future work.

The paper is organized as follows.
In sect. \ref{sec:LAA}, we review briefly 
the basic notions of the theory of Lie algebroids and 
Lie algebroid infinitesimal actions on fibered Poisson manifolds.
In sect. \ref{sec:LAsigma}, we illustrate the Lie algebroid Poisson sigma model and its main properties.
In sect. \ref{sec:LAH*},  we review briefly various cohomologies associated to the target 
space geometry of the model. 
In sect. \ref{sec:LAcohom},  relying on the results of sect. \ref{sec:LAH*}, 
we study in detail the BV cohomology.
In sect. \ref{sec:LGG},  after reviewing briefly the basic notions of the theory of Lie groupoids,
we describe the generalized fiberwise Hamiltonian reduction encoded in the target space geometry 
of the model in the integrable case.
In sect. \ref{sec:exa}, we illustrate a few examples.
Sect. \ref{sec:conclusions} contains some concluding remarks.
Finally, in app. \ref{app:covariance}--\ref{app:diff}, 
we conveniently collect some of the technical details of the 
analysis expounded in the main body of the paper.

\vskip.5cm
\textcolor{blue}{\bf Acknowledgments}. We thank Y. Kosmann--Schwarzbach, M. Crainic and D. Roytenberg 
for sharing with us their insight on Lie algebroid theory. 

\vfill\eject

\section{\normalsize \textcolor{blue}{Lie algebroids and their 
action on Poisson manifolds}}\label{sec:LAA}

\par
\hskip.6cm In this section, we shall review briefly 
the basic notions of the theory of Lie algebroids and 
Lie algebroid infinitesimal actions \cite{Mackenzie1,Cartier1,Mackenzie2}, concentrating 
on actions on fibered Poisson manifolds
because of their relevance on the following constructions \cite{Vaisman}.
We have expressed the relevant geometrical relations also in local coordinates
to help the reader to check the calculation on his/her own.

A {\it Lie algebroid} is a vector bundle $L$ over a manifold $M$ 
equipped with a bundle map $\rho:L\rightarrow TM$, called the {\it anchor} and 
an $\mathbb{R}$--linear bracket $[\cdot,\cdot]:\Gamma(L)\times \Gamma(L)
\rightarrow \Gamma(L)$ with the following properties.

\par\noindent 
~~~1) $[\cdot,\cdot]$ is a Lie bracket so that $\Gamma(L)$ is a Lie algebra:
\begin{align}
&[s,t]+[t,s]=0,
\vphantom{\bigg]}
\label{LAA1}
\\
&[s,[t,u]]+[t,[u,s]]+[u,[s,t]]=0,
\vphantom{\bigg]}
\label{LAA2}
\end{align}
for $s,t,u\in \Gamma(L)$. 

\par\noindent 
~~~2) $\rho$ defines a Lie algebra morphism of $\Gamma(L)$ into $\Gamma(TM)$: 
\begin{equation}\label{LAA3}
\rho([s,t])=[\rho(s),\rho(t)]_{TM},
\end{equation}
for $s,t\in \Gamma(L)$, where $[\cdot,\cdot]_{TM}$ is the usual Lie bracket
of vector fields of $M$. 

\par\noindent 
~~~3) The Leibniz rule holds:
\begin{equation}\label{LAA4}
[s,ft]=f[s,t]+(l_{\rho(s)}f)t,
\end{equation}
for $f\in C^\infty(M)$ and $s,t\in \Gamma(L)$, where $l_v$ denotes Lie derivation along a vector
field $v\in \Gamma(TM)$. 

The prototype Lie algebroid over $M$ is the tangent bundle $TM$: the anchor is the
identity $\id_{TM}$ and the bracket is the usual Lie bracket $[\cdot,\cdot]_{TM}$. 
Lie algebroids generalize Lie algebras: a Lie algebra can be viewed as 
a Lie algebroid over the singleton manifold $M=\mathrm{pt}$.

Let $\{e_i\}$ be a local frame of $L$. Then, one has
\begin{align}
&\rho(e_i)=\rho_i{}^r\partial_r,
\vphantom{\bigg]}
\label{LAA5}
\\
&[e_i,e_j]=f^k{}_{ij}e_k.
\vphantom{\bigg]}
\label{LAA6}
\end{align}
$\rho_i{}^r$, $f^k{}_{ij}$ are called the anchor and structure functions of $L$, respectively.
From \eqref{LAA1}--\eqref{LAA4}, they satisfy
\begin{align}
&f^i{}_{jk}+f^i{}_{kj}=0,
\vphantom{\bigg]}
\label{LAA7}
\\
&f^i{}_{jm}f^m{}_{kl}+f^i{}_{km}f^m{}_{lj}+f^i{}_{lm}f^m{}_{jk}
+\rho_j{}^r\partial_rf^i{}_{kl}+\rho_k{}^r\partial_rf^i{}_{lj}
+\rho_l{}^r\partial_rf^i{}_{jk}=0,
\vphantom{\bigg]}
\label{LAA8}
\\
&\rho_i{}^s\partial_s\rho_j{}^r-\rho_j{}^s\partial_s\rho_i{}^r-f^k{}_{ij}\rho_k{}^r=0.
\vphantom{\bigg]}
\label{LAA9}
\end{align}

A Lie algebroid $L$ over $M$ is said {\it regular}, if the anchor $\rho$ has locally constant rank.
In such a case, $\ker\rho$ is a bundle of Lie algebras, but not a Lie algebra bundle in 
general. If $L$ is regular, $\Gamma(\ker\rho)$ is a Lie ideal of $\Gamma(L)$.
A Lie algebroid is said {\it transitive}, if $\rho$ is surjective. A transitive Lie algebroid is 
obviously regular and its $\ker\rho$ is a Lie algebra bundle.

For a regular Lie algebroid, we can choose adapted frames $\{e_i\}=\{e_\alpha\}\cup\{e_\kappa\}$,
where $\{e_\alpha\}$ is a frame of $\ker\rho$. Such frames will be tacitly assumed, unless otherwise stated.
Clearly, one has 
\begin{equation}\label{LAA10}
\rho_\alpha{}^r=0
\end{equation}
identically. Further, \hphantom{xxxxxxxxxxxxxxxxxxxxxx}
\begin{equation}\label{LAA11}
f^\kappa{}_{i\alpha}=0,
\end{equation}
since $\Gamma(\ker\rho)$ is a Lie ideal.

A {\it (base preserving) morphism} of two Lie algebroids $L$, $L'$ over $M$ is a vector bundle 
morphism $\varphi:L\rightarrow L'$ such that
\vfill\eject
\begin{align}
&\rho'\circ\varphi=\rho,
\vphantom{\bigg]}
\label{LAAmorph1}
\\
&\varphi([s,t])=[\varphi(s),\varphi(t)]',
\vphantom{\bigg]}
\label{LAAmorph2}
\end{align}
with $s,t\in \Gamma(L)$. 

If $K$, $L$ are two Lie algebroids over $M$ and $K$ is a subbundle 
of $L$, then $K$ is a {\it subalgebroid} of $L$, if the natural injection
$\iota:K\rightarrow L$ is a Lie algebroid morphism. 
If $L$ is regular, then $K=\ker\rho$ is a subalgebroid of $L$.

We recall that a {\it fibered manifold} is a manifold $X$ together with 
a surjective submersion $J:X\rightarrow M$ onto another manifold $M$.  

Let $L$ be a Lie algebroid over $M$ and let $J:X\rightarrow M$ be a fibered manifold.
An {\it infinitesimal action} of $L$ on $X$ along $J$ is an $\mathbb{R}$--linear map
$u:\Gamma(L)\rightarrow \Gamma(TX)$ with the following properties \cite{Mackenzie2}.

\par\noindent 
~~~1) $u$ is $C^\infty(M)$--linear: \hphantom{xxxxxxxxxxxxxxxxxxxxxx}
\begin{equation}\label{LAA12}
u(fs)=(f\circ J)u(s),
\end{equation}
for $f\in C^\infty(M)$ and $s\in \Gamma(L)$.

\par\noindent 
~~~2) $u$ is a Lie algebra morphism:
\begin{equation}\label{LAA13}
u([s,t])=[u(s),u(t)]_{TX},
\end{equation}
for $s,t\in \Gamma(L)$. 

\par\noindent 
~~~3) $u$ is projectable: \hphantom{xxxxxxxxxxxxxxxxxxxxxx}
\begin{equation}\label{LAA14}
TJ(u(s))=\rho(s)\circ J,
\end{equation}
for $s\in \Gamma(L)$, where $TJ$ is the tangent map of $J$. 

\par\noindent 
The last conditions implies that the vector fields $u(s)$ with $s\in \Gamma(\ker\rho)$ 
are tangent to the fibers of $J$.

Upon picking a frame $\{e_i\}$ of $L$, one has
\begin{equation}\label{LAA15}
u(e_i)=u_i{}^a\partial_a,
\end{equation}
where $u_i{}^a$ are the Lie algebroid action functions. From \eqref{LAA12}--\eqref{LAA14}, they satisfy: 
\begin{align}
&u_i{}^b\partial_bu_j{}^a-u_j{}^b\partial_bu_i{}^a-f^k{}_{ij}\circ Ju_k{}^a=0,
\vphantom{\bigg]}
\label{LAA16}
\\
&u_i{}^a\partial_aJ^r=\rho_i{}^r\circ J.
\vphantom{\bigg]}
\label{LAA17}
\end{align}

To an infinitesimal action of $L$ on $X$ along $J$, there is canonically associated a Lie algebroid 
structure on the pull back vector bundle $J^*L$. Its definition invokes the isomorphism
$C^\infty(X)\otimes_{C^\infty(M)}\Gamma(L)\simeq \Gamma(J^*L)$ given by
$f\otimes s\rightarrow f(s\circ J)$ with $f\in C^\infty(X)$, $s\in \Gamma(L)$. 
The anchor and Lie bracket are 
then defined by
\begin{align}
&\rho_J(f\otimes s)=fu(s),
\vphantom{\bigg]}
\label{LAAaLa1}
\\
&[f\otimes s,g\otimes t]_J=fg\otimes [s,t]+(fl_{u(s)}g)\otimes t-(gl_{u(t)}f)\otimes s,
\vphantom{\bigg]}
\label{LAAaLa2}
\end{align}
for $f, g\in C^\infty(X)$, $s,t\in \Gamma(L)$.
The resulting Lie algebroid is called the {\it action Lie algebroid}
corresponding to the infinitesimal action and is usually denoted by 
$L\ltimes J$. The anchor and structure functions of $L\ltimes J$ are $u_i{}^a$ and $f^i{}_{jk}\circ J$, 
respectively, as is easy to see.

A {\it Poisson structure} on a manifold $X$ is a 2--vector field $P\in \Gamma(\wedge^2TX)$
satisfying the Poisson condition 
\begin{equation}\label{LAA18}
[P,P]_{\wedge^*TX}=0,
\end{equation}
where $[\cdot,\cdot]_{\wedge^*TX}$ is the Schouten--Nijenhuis bracket.
The Poisson structure $P$ allows the definition of a Poisson bracket on $X$
by setting $\{f,g\}_P=P(df,dg)$, where $f,g\in C^\infty(X)$. $X$ is thus called a 
{\it Poisson manifold}. 

In local coordinates, $P$ is given by 
\begin{equation}\label{LAA20}
P=\frac{1}{2}P^{ab}\partial_a\wedge\partial_b.
\end{equation}
Then, the Poisson condition \eqref{LAA18} reads
\begin{equation}\label{LAA21}
P^{ad}\partial_dP^{bc}+P^{bd}\partial_dP^{ca}+P^{cd}\partial_dP^{ab}=0.
\end{equation}
 
A {\it fibered Poisson manifold} is a fibered manifold 
$J:X\rightarrow M$ together with a Poisson structure $P$ on $X$ satisfying the condition
\begin{equation}\label{LAA22}
P(TJ)^*=0,
\end{equation}
where we view $P\in \Gamma(\Hom(T^*X,TX))$. Intuitively, this means that the 2--vector field
$P$ is everywhere tangent to the fibers of $J$. 
$X$ can then be viewed as a family of Poisson manifolds smoothly parametrized by $M$. 

In local coordinates, \eqref{LAA22} reads simply
\begin{equation}\label{LAA23}
P^{ab}\partial_b J^r=0.
\end{equation}

Let $L$ be a Lie algebroid over $M$ and $J:X\rightarrow M,P$ be a fibered Poisson manifold
and let $L$ act infinitesimally on $X$ along $J$. $P$ is said {\it invariant} if
\begin{equation}\label{LAA24}
l_{u(s)}P=0,
\end{equation}
for $s\in \Gamma(L)$.

In local coordinates, the invariance condition \eqref{LAA24} reads 
\begin{equation}\label{LAA25}
u_i{}^c\partial_c P^{ab}-\partial_cu_i{}^aP^{cb}-\partial_cu_i{}^bP^{ac}=0.
\end{equation}

Let $L$ be a regular Lie algebroid over $M$ and $J:X\rightarrow M,P$ be a fibered Poisson manifold
and let $L$ act infinitesimally on $X$ along $J$ leaving $P$ invariant.
An {\it equivariant moment map} for the action
is an $\mathbb{R}$--linear map $\mu:\Gamma(\ker \rho)\rightarrow C^\infty(X)$ 
with the following properties \cite{Marsden1,Ratiu1,Bos1}.

\par\noindent 
~~~1) $\mu$ is $C^\infty(M)$--linear: \hphantom{xxxxxxxxxxxxxxxxxxxxxx}
\begin{equation}\label{LAA26}
\mu(fs)=(f\circ J)\mu(s),
\end{equation}
for $f\in C^\infty(M)$ and $s\in \Gamma(\ker\rho)$.

\par\noindent 
~~~2) $\mu$ is equivariant: \hphantom{xxxxxxxxxxxxxxxxxxxxxx}
\begin{equation}\label{LAA27}
l_{u(s)}\mu(t)=\mu([s,t]),
\end{equation}
for $s\in \Gamma(L)$ and $t\in \Gamma(\ker\rho)$.

\par\noindent 
~~~3) $\mu$ is a moment map for $u$: \hphantom{xxxxxxxxxxxxxxxxxxxxxx} 
\begin{equation}\label{LAA28}
u(s)=\#_Pd_X\mu(s),
\end{equation}
for $s\in \Gamma(\ker\rho)$, where $\#_P:T^*X\rightarrow TX$ is the sharp map associated to $P$
by viewing $P\in \Gamma(\Hom(T^*X,TX))$. These relations imply that 
\begin{equation}\label{LAA29}
\{\mu(s),\mu(t)\}_P=\mu([s,t]),
\end{equation}
for $s,t\in  \Gamma(\ker\rho)$.

Pick an adapted frame $\{e_i\}$ of $L$ and set
\begin{equation}\label{LAA19}
\mu_\alpha=\mu(e_\alpha).
\end{equation}
Then,\eqref{LAA27}, \eqref{LAA28} in local coordinates read
\begin{align}
&u_\alpha{}^a+P^{ab}\partial_b\mu_\alpha=0,
\vphantom{\bigg]}
\label{LAA30}
\\
&u_i{}^a\partial_a\mu_\alpha-f^\beta{}_{i\alpha}\circ J\mu_\beta=0.
\vphantom{\bigg]}
\label{LAA31}
\end{align}

The fact that 
$\mu(s)$ is defined for $s\in \Gamma(\ker\rho)$ rather than  $s\in \Gamma(L)$
may seem puzzling at first glance. In regard to this, 
let us note that, if $\mu(s)$ were defined for $s\in \Gamma(L)$,
the equivariance condition \eqref{LAA27} would not be covariant. Let us note further that, 
when $L$ is an ordinary Lie algebra $\mathfrak{g}$, then $\ker\rho=L$ and one has 
$\mu:\mathfrak{g}\rightarrow C^\infty(X)$ as usual.

In the geometrical framework illustrated above, a pivotal role is played by
the action of a Lie algebroid $L$ over $M$ on 
a fibered manifold $J:X\rightarrow M,P$. In this respect,  
there are two important extremal cases deserving mention.

\par\noindent 
~~~$a$) $M=X$ and $J:X\rightarrow X$ the identity map. In this case,
the infinitesimal action of $L$ on $X$ reduces to the canonical infinitesimal 
action of $L$ on $M$, for which $u=\rho$. Further, $P$ is necessarily trivial 
and thus trivially invariant under $L$  (see eqs. \eqref{LAA22}, \eqref{LAA24}).


\par\noindent 
~~~$b$) $M=\mathrm{pt}$ and $J:X\rightarrow\mathrm{pt}$ the constant map.
In this case, the infinitesimal action of $L$ on $X$
yields an ordinary infinitesimal action of the Lie algebra associated to $L$ 
on $X$, whose fundamental vector field is $u$. 
Further, the Poisson structure $P$ is subject only to the invariance condition 
under the Lie algebra action but is otherwise arbitrary
(see again eqs. \eqref{LAA22}, \eqref{LAA24}).


The general case is in a sense intermediate and interpolates between the two above extremal cases.

\vfill\eject

\section{\normalsize \textcolor{blue}{The Lie algebroid Poisson sigma model}}\label{sec:LAsigma}

\par
\hskip.6cm In this section, we shall construct a sigma model canonically associated to the following 
geometrical data (cf. sect. \ref{sec:LAA}).
\begin{enumerate}

\item A regular Lie algebroid $L$ over $M$.

\item A fibered Poisson manifold $J:X\rightarrow M,P$

\item An infinitesimal action of $L$ on $X$ along $J$ leaving $P$ invariant.

\item A equivariant moment map $\mu$ for the action.
\end{enumerate}
We shall call it {\it Lie algebroid Poisson sigma model} for evident reasons. We shall use a BV formalism
\cite{BV1,BV2} following the geometrical approach of AKSZ \cite{AKSZ} and Cattaneo and Felder
\cite{Cattaneo1,Cattaneo2}. 

The base space of the model is the parity shifted tangent bundle $T[1]\Sigma$
of a closed surface $\Sigma$, the world sheet.
The target space of the model
is a graded manifold, the parity shifted vector bundle over $X$ 
\begin{equation}\label{LAsigma1}
\mathfrak{X}_{L,J}=(J^*L)^*[0]\oplus(J^*\ker\rho)^*[-1].
\end{equation}
The fields of the model organize in a superfield
$\mathbfs{\Phi}\in C^\infty(T[1]\Sigma,T^*[1]\mathfrak{X}_{L,J})$, where $T^*[1]\mathfrak{X}_{L,J}$
is the parity shifted cotangent bundle of $\mathfrak{X}_{L,J}$.
Locally in target space, $\mathbfs{\Phi}$ is given as a sextuplet of superfields
$(\mathbfs{x}^a,\mathbfs{b}_i,\mathbfs{B}_\alpha,\mathbfs{y}_a,\mathbfs{c}^i,\mathbfs{C}^\alpha)$
of degrees $(0,0,-1,1,1,2)$, respectively. The triples $(\mathbfs{x}^a,\mathbfs{b}_i,\mathbfs{B}_\alpha)$,  
$(\mathbfs{y}_a,\mathbfs{c}^i,\mathbfs{C}^\alpha)$ correspond to the base and fiber coordinates of
$T^*[1]\mathfrak{X}_{L,J}$, respectively. In turn, $\mathbfs{x}^a$, $(\mathbfs{b}_i,\mathbfs{B}_\alpha)$ 
correspond to the base and fiber coordinates of 
$\mathfrak{X}_{L,J}$.
We note that $\mathbfs{x}\in C^\infty(T[1]\Sigma,X)$ and that 
$\mathbfs{b}\in \Gamma(\mathbfs{x}^*((J^*L)^*[0]))$,
$\mathbfs{B}\in \Gamma(\mathbfs{x}^*((J^*\ker \rho)^*[-1]))$,
$\mathbfs{c}\in \Gamma(\mathbfs{x}^*(J^*L[1]))$,
$\mathbfs{C}\in \Gamma(\mathbfs{x}^*(J^*\ker\rho[2]))$, while $\mathbfs{y}$ does 
not have an analogous simple interpretation.
See app. \ref{app:covariance} for details
on covariance for the manifold $T^*[1]\mathfrak{X}_{L,J}$. 

The field space is equipped with a degree $-1$ symplectic form obtained by pulling back 
with the evaluation map of  
$C^\infty(T[1]\Sigma,T^*[1]\mathfrak{X}_{L,J})$ the canonical symplectic form of $T^*[1]\mathfrak{X}_{L,J}$ 
and then integrating over $T[1]\Sigma$:
\begin{equation}\label{LAsigma2}
\Omega_{L,J}=\int_{T[1]\Sigma}\varrho\Big[\delta\mathbfs{x}^a\delta\mathbfs{y}_a
+\delta\mathbfs{b}_i\delta\mathbfs{c}^i
+\delta\mathbfs{B}_\alpha\delta\mathbfs{C}^\alpha\Big],
\end{equation}
where $\varrho$ is the invariant supermeasure on $T[1]\Sigma$.
From this, one obtains the BV antibracket $(\cdot,\cdot)_{L,X}$ in standard fashion:
\begin{align}
(F,G)_{L,J}=\int_{T[1]\Sigma}\varrho\bigg[\frac{\delta_rF}{\delta\mathbfs{x}^a}\frac{\delta_lG}{\delta\mathbfs{y}_a}
-\frac{\delta_rF}{\delta\mathbfs{y}_a}&\frac{\delta_lG}{\delta\mathbfs{x}^a}
+\frac{\delta_rF}{\delta\mathbfs{b}_i}\frac{\delta_lG}{\delta\mathbfs{c}^i}
\vphantom{\int_\Sigma}
\label{LAsigma3}
\\
&-\frac{\delta_rF}{\delta\mathbfs{c}^i}\frac{\delta_lG}{\delta\mathbfs{b}_i}
+\frac{\delta_rF}{\delta\mathbfs{B}_\alpha}\frac{\delta_lG}{\delta\mathbfs{C}^\alpha}
-\frac{\delta_rF}{\delta\mathbfs{C}^\alpha}\frac{\delta_lG}{\delta\mathbfs{B}_\alpha}\bigg],
\vphantom{\int_\Sigma}
\nonumber
\end{align}
where $\delta_{l,r}/\delta\mathbfs{\phi}$ denotes left/right functional derivation with respect to the superfield
$\mathbfs{\phi}$.

The action of the model is
\begin{align}
S_{J,P}&=\int_{T[1]\Sigma}\varrho\Big[
\mathbfs{y}_a\big(\mathbfs{d}\mathbfs{x}^a+u_i{}^a(\mathbfs{x})\mathbfs{c}^i\big)
+\mu_\alpha(\mathbfs{x})\mathbfs{C}^\alpha
+\frac{1}{2}P^{ab}(\mathbfs{x})\mathbfs{y}_a\mathbfs{y}_b
\vphantom{\int_\Sigma}
\label{LAsigma4}
\\
&\hskip0.9cm+\mathbfs{b}_i\big(\mathbfs{d}\mathbfs{c}^i
-\frac{1}{2}f^i{}_{jk}(J(\mathbfs{x}))\mathbfs{c}^j\mathbfs{c}^k+\delta^i{}_\alpha\mathbfs{C}^\alpha\big)
-\mathbfs{B}_\alpha\big(\mathbfs{d}\mathbfs{C}^\alpha-f^\alpha{}_{i\beta}(J(\mathbfs{x}))
\mathbfs{c}^i\mathbfs{C}^\beta\big)\Big].
\nonumber
\end{align}
For the target space global definedness of the integrand, it is absolutely crucial that 
the Poisson structure $P$ satisfies the tangentiality condition \eqref{LAA23}.
This follows straightforwardly from eqs. \eqref{chcoord*} in app. \ref{app:covariance}.
Above, one views $u\in \Gamma(\Hom(J^*L,TX))$ and $\mu\in \Gamma((J^*\ker\rho)^*)$, 
as allowed by \eqref{LAA12}, \eqref{LAA26}.

The properties of the target space geometry of the sigma model make $S_{J,P}$ satisfy the BV classical 
master equation \cite{BV1,BV2} \hphantom{xxxxxxxxxxxxxxxxxxxxx}
\begin{equation}\label{LAsigma5}
(S_{J,P},S_{J,P})_{L,J}=0.
\end{equation}
The verification is a straightforward calculation exploiting certain combinations of the local coordinate relations
\eqref{LAA7},  \eqref{LAA8}, \eqref{LAA10}, \eqref{LAA11}, \eqref{LAA16}, \eqref{LAA17}, \eqref{LAA21}, 
\eqref{LAA23}, \eqref{LAA25}, \eqref{LAA30}, \eqref{LAA31}.
We observe that these relations are sufficient but not necessary conditions for the validity
of \eqref{LAsigma5}. This fact is a recurrent feature of the AKSZ formulation of sigma models. 

Associated with the master action $S_{J,P}$ is the BV field variation operator $\delta_{J,P}:=(S_{J,P},\cdot)_{L,J}$.
The BV field variations are: 
\begin{subequations}\label{LAsigma6}
\begin{align}
&\delta_{J,P}\mathbfs{x}^a=\mathbfs{d}\mathbfs{x}^a+u_i{}^a(\mathbfs{x})\mathbfs{c}^i+P^{ab}(\mathbfs{x})\mathbfs{y}_b,
\vphantom{\frac{1}{2}}
\label{LAsigma6a}
\\
&\delta_{J,P}\mathbfs{y}_a=\mathbfs{d}\mathbfs{y}_a+\partial_au_i{}^b(\mathbfs{x})\mathbfs{y}_b\mathbfs{c}^i
+\partial_a\mu_\alpha(\mathbfs{x})\mathbfs{C}^\alpha
+\frac{1}{2}\partial_aP^{bc}(\mathbfs{x})\mathbfs{y}_b\mathbfs{y}_c
\vphantom{\frac{1}{2}}
\label{LAsigma6b}
\\
&\hskip.5cm-\frac{1}{2}\partial_aJ^r(\mathbfs{x})\partial_rf^i{}_{jk}(J(\mathbfs{x}))\mathbfs{b}_i\mathbfs{c}^j\mathbfs{c}^k
-\partial_aJ^r(\mathbfs{x})\partial_rf^\alpha{}_{i\beta}(J(\mathbfs{x}))\mathbfs{c}^i\mathbfs{B}_\alpha\mathbfs{C}^\beta,
\vphantom{\frac{1}{2}}
\nonumber
\\
&\delta_{J,P}\mathbfs{c}^i=\mathbfs{d}\mathbfs{c}^i-\frac{1}{2}f^i{}_{jk}(J(\mathbfs{x}))\mathbfs{c}^j\mathbfs{c}^k
+\delta^i{}_\alpha\mathbfs{C}^\alpha,
\vphantom{\frac{1}{2}}
\label{LAsigma6c}
\\
&\delta_{J,P}\mathbfs{b}_i=\mathbfs{d}\mathbfs{b}_i+f^j{}_{ki}(J(\mathbfs{x}))\mathbfs{b}_j\mathbfs{c}^k
+f^\alpha{}_{\beta i}(J(\mathbfs{x}))\mathbfs{B}_\alpha\mathbfs{C}^\beta-u_i{}^a(\mathbfs{x})\mathbfs{y}_a,
\vphantom{\frac{1}{2}}
\label{LAsigma6d}
\\
&\delta_{J,P}\mathbfs{C}^\alpha=\mathbfs{d}\mathbfs{C}^\alpha-f^\alpha{}_{i\beta}(J(\mathbfs{x}))\mathbfs{c}^i\mathbfs{C}^\beta,
\vphantom{\frac{1}{2}}
\label{LAsigma6e}
\\
&\delta_{J,P}\mathbfs{B}_\alpha=\mathbfs{d}\mathbfs{B}_\alpha+f^\beta{}_{i\alpha}(J(\mathbfs{x}))\mathbfs{c}^i\mathbfs{B}_\beta
-\mathbfs{b}_\alpha-\mu_\alpha(\mathbfs{x}). \hskip2cm 
\vphantom{\frac{1}{2}}
\label{LAsigma6f}
\end{align} 
\end{subequations}
The master equation \eqref{LAsigma5} implies that $S_{J,P}$ is invariant under $\delta_{J,P}$, 
\begin{equation}\label{LAsigma7}
\delta_{J,P}S_{J,P}=0
\end{equation}
and that $\delta_{J,P}$ is nilpotent \hphantom{xxxxxxxxxxxxxxxxxxxxx}
\begin{equation}\label{LAsigma8}
\delta_{J,P}{}^2=0.
\end{equation}

It is interesting to examine what happens in the two extremal cases considered at the end of sect. 
\ref{sec:LAA}.

\par\noindent 
~~~$a$) In this case, $M=X$ and $J$ is the identity map $\id_X$. Since $P=0$ identically,  
if we also set $\mu=0$, we get a sigma model canonically associated to the Lie algebroid $L$,
which we call {\it Lie algebroid sigma model}.
It is simple to check that  
the basic relations obeyed by the anchor and the structure functions,
eqs. \eqref{LAA8}, \eqref{LAA9}, are not only sufficient but also necessary for the BV classical 
master equation \eqref{LAsigma5} to hold.

\par\noindent 
~~~$b$) In this case, $M=\mathrm{pt}$ and $J$ is the constant map.
$L$ is an ordinary Lie algebra acting infinitesimally on $X$ and 
$P$ is invariant under such an action. The resulting sigma model
is nothing but the Poisson--Weil model of refs.\cite{Zucchini6,Zucchini7} 
for trivial twisting principal bundle. See sect. \ref{sec:conclusions}
for more on this point. 

For general $M$ and $J$, the Lie algebroid Poisson sigma model
is consistently defined for $P=0$ and $\mu=0$ provided
$u(s)=0$ for $s\in \Gamma(\ker\rho)$. (In general, only the weaker condition
$TJ(u(s))=0$ holds, see eq. \eqref{LAA14}.)
In that case, it reduces into the Lie algebroid sigma model
of the action Lie algebroid $L\ltimes J$
(cf. sect. \ref{sec:LAA}). 

Inspection of the action \eqref{LAsigma4} reveals that the Lie algebroid sigma model
is a Poisson sigma model on the graded manifold $\mathfrak{X}_{L,J}$ twisted 
by a moment map potential term. 
The target space 2--vector $\Pi\in \Gamma(\wedge^2T\mathfrak{X}_{L,J})$ 
of this Poisson sigma model is given by the following expressions:
\begin{subequations}\label{LAsigma9}
\begin{align}
&\Pi^{ab}(\Xi)=P^{ab}(\xi),
\vphantom{\bigg]}
\label{LAsigma9a}
\\
&\Pi^a{}_i(\Xi)=u_i{}^a(\xi),
\vphantom{\bigg]}
\label{LAsigma9b}
\\
&\Pi^a{}_\alpha(\Xi)=0,
\vphantom{\bigg]}
\label{LAsigma9c}
\\
&\Pi_{ij}(\Xi)=-f^k{}_{ij}(J(\xi))\beta_k,
\vphantom{\bigg]}
\label{LAsigma9d}
\\
&\Pi_{i\alpha}(\Xi)=-f^\beta{}_{i\alpha}(J(\xi))\Beta_\beta,
\vphantom{\bigg]}
\label{LAsigma9e}
\\
&\Pi_{\alpha\beta}(\Xi)=0,
\vphantom{\bigg]}
\label{LAsigma9f}
\end{align}
\end{subequations}
where $\xi^a$, $(\beta_i,\Beta_\alpha)$ are respectively the base and fiber coordinates of the bundle
$\mathfrak{X}_{L,J}=(J^*L)^*[0]\oplus(J^*\ker\rho)^*[-1]$ 
and we have set $\Xi^A=(\xi^a,\beta_i,\Beta_\alpha)$ (cf. app. \ref{app:covariance}).
The relations \eqref{LAA7},  \eqref{LAA8}, \eqref{LAA10}, \eqref{LAA11}, \eqref{LAA16}, 
\eqref{LAA17}, \eqref{LAA21}, \eqref{LAA23}, \eqref{LAA25}, 
which ensure the fulfillment of the BV 
classical master equation \eqref{LAsigma5}, also ensure that the 2--vector $\Pi$ satisfies the 
Poisson condition.

\vfill\eject

\section{\normalsize \textcolor{blue}{Action Lie algebroid and Poisson cohomology}}\label{sec:LAH*}

\par
\hskip.6cm The Lie algebroid Poisson sigma model introduced in sect. \ref{sec:LAsigma}
is characterized by the associated BV cohomology \cite{Mackenzie1,Cartier1}. 
This in turn is intimately related with various cohomologies associated with the target 
space geometry. In this section, we briefly review them.  

A Lie algebroid $L$ over $M$ is endowed with a natural cohomology,
the {\it Lie algebroid cohomology}. 
This is the cohomology of the complex  $(A^{*}(L), d_L)$, where $A^p(L)=\Gamma(\wedge^pL^*)$ 
consists of $C^{\infty}(M)$--multilinear antisymmetric maps $\omega:\Gamma(L)^p\rightarrow C^\infty(M)$ and  
the nilpotent differential $d_L:A^p(L)\rightarrow A^{p+1}(L)$ is given by the well--known Che\-val\-ley-Eilenberg formula:
\begin{align}
(d_L\omega)(s_1,\ldots, s_{p+1}) 
&=\sum_i(-1)^{i+1}l_{\rho(s_i)}(\omega(s_1,\ldots,\hat s_i,\ldots,s_{p+1})) 
\vphantom{\bigg[}
\label{LAAH*1}
\\
&\,+\sum_{i<j}(-1)^{i+j}\omega([s_i, s_j], s_1, \ldots , \hat s_i, \ldots,\hat s_j, \ldots s_{p+1}),
\nonumber
\end{align}
with $s_1,\ldots,s_{p+1}\in \Gamma(L)$. 

The Lie algebroid cohomology complex can be described alternatively in supergeometric terms as follows.
There is an 
isomorphism $\Gamma(\wedge^*L^*)\simeq C^\infty(L[1])$ defined by 
$\omega\rightarrow \frac{1}{p!}\omega(\xi,\ldots,\xi)$ with $\omega\in \Gamma(\wedge^pL^*)$, where
$\xi=\xi^i\otimes e_i$, $e_i$ and $\xi^i$ being the elements of a local frame of $L$ and 
the corresponding degree $1$ fiber coordinates of $L[1]$, respectively. 
Under the isomorphism, the differential $d_L$ turns into the homological vector field over $L[1]$
given by 
\begin{equation}\label{LAAH*1'}
d_L=
\xi^il_{\rho(e_i)}-\frac{1}{2}f^k{}_{ij}\xi^i\xi^j\partial_{\xi k},
\end{equation}
where $\partial_{\xi i}=\partial/\partial \xi^i$.
The supergeometric formulation is more convenient in general.

A {\it representation} of a Lie algebroid $L$ over $M$ is a vector bundle $E$ over $M$
together with an assignment to each $s\in \Gamma(L)$ of an $\mathbb{R}$--linear map
$D_s:\Gamma(E)\rightarrow \Gamma(E)$ with the following properties:
\vfill\eject
\begin{subequations}\label{LAAH*2}
\begin{align}
&D_{fs}\sigma=fD_s\sigma,
\vphantom{\bigg[}
\label{LAAH*2a}
\\
&D_s(f\sigma)=fD_s\sigma+(l_{\rho(s)}f)\sigma,
\vphantom{\bigg[}
\label{LAAH*2b}
\\
&[D_s,D_t]\sigma=D_{[s,t]}\sigma,
\vphantom{\bigg[}
\label{LAAH*2c}
\end{align}
\end{subequations}
where $s,t\in \Gamma(L)$, $f\in C^\infty(M)$ and $\sigma\in \Gamma(E)$.
The trivial representations is defined by $E=M\times\mathbb{R}$ and
$D_sh=l_{\rho(s)}h$, with $s\in \Gamma(L)$ and $h\in C^\infty(M)$.
If $L$ is regular (cf. sect. \ref{sec:LAA}), the adjoint representation is defined by
$E=\ker\rho$ and $D_su=[s,u]$, with $s\in \Gamma(L)$ and $u\in \Gamma(\ker\rho)$.
If the base $M$ is a point, then $L$ is a Lie algebra and $E$ is a vector space, and 
a representation of $L$ is just an ordinary Lie algebra linear representation. 

One can define the Lie algebroid cohomology of $L$ with values in a given representation $D$ 
of $L$. This is the cohomology of the complex $(A^{*}(L,D), d_{L,D})$, where $A^p(L,D)=\Gamma(\wedge^pL^*\otimes E)$ 
consists of $C^{\infty}(M)$--multilinear antisymmetric maps $\omega:\Gamma(L)^p\rightarrow \Gamma(E)$ and  
the nilpotent differential $d_{L,D}:A^p(L,D)\rightarrow A^{p+1}(L,D)$ is given by the 
Chevalley-Eilenberg formula \eqref{LAAH*1} with $l_{\rho(s_i)}$ replaced by $D_{s_i}$.
There is also a supergeometric formulation exploiting the isomorphism
$\Gamma(\wedge^*L^*\otimes E)\simeq \Gamma(\pi_{L[1]}{}^*E)$, 
where $\pi_{L[1]}:L[1]\rightarrow M$ is the bundle projection, 
in which $d_{L,D}$ turns into a 
homological vector field over $L[1]$ given by
\eqref{LAAH*1'} with $l_{\rho(e_i)}$ replaced by $D_{e_i}$. When $E=M\times\mathbb{R}$ 
and $D_s=l_{\rho(s)}$, one recovers the usual Lie algebroid cohomology. 

Let a Lie algebroid $L$ over $M$ act infinitesimally on a fibered manifold 
$J:X\rightarrow M$  (cf. sect. \ref{sec:LAA}).
To the action, one can associate the {\it action Lie algebroid cohomology},
the Lie algebroid cohomology of the action Lie algebroid $L\ltimes J$ (cf. sect. \ref{sec:LAA}).
The associated cochain complex can be described as follows. The cochain space 
$A^p(L\ltimes J)$ consists of the antisymmetric maps $\omega:\Gamma(L)^p\rightarrow C^\infty(X)$ 
which are $C^{\infty}(M)$--multilinear meaning that
\begin{equation}
\label{LAAH*4}
\omega(s_1,\ldots,fs_m,\ldots s_p)=(f\circ J)\omega(s_1,\ldots,s_m,\ldots s_p),
\end{equation}
with $s_1,\ldots,s_p\in \Gamma(L)$ and $f\in C^\infty(M)$. The differential $d_{L\ltimes J}$ 
is then given by the Chevalley-Eilenberg formula \eqref{LAAH*1} with $\rho(s_i)$ substituted by $u(s_i)$.
In the supergeometric formulation,  $d_{L\ltimes J}$  is given by \eqref{LAAH*1'} 
with $l_{\rho(e_i)}$ replaced by $l_{u(e_i)}$ and $f^k{}_{ij}$ by $f^k{}_{ij}\circ J$.

For analogous reasons, a representation $D$ of $L\ltimes J$ on a vector bundle $E$ 
over $X$ can be described as an assignment to each $s\in \Gamma(L)$ of an $\mathbb{R}$--linear map
$D_s:\Gamma(E)\rightarrow \Gamma(E)$ satisfying
\begin{equation}\label{LAAH*5}
D_{fs}\sigma=(f\circ J)D_s\sigma,
\end{equation}
for $s\in \Gamma(L)$, $f\in C^\infty(M)$ and $\sigma\in \Gamma(E)$,
in substitution of \eqref{LAAH*2a}, together with \eqref{LAAH*2b}, \eqref{LAAH*2c} 
with $s,t\in \Gamma(L)$, $f\in C^\infty(X)$ and $\sigma\in \Gamma(E)$
and $\rho(s)$ replaced by $u(s)$. The action Lie algebroid cohomology with values in $E$
can then be described as follows.
The cochain space 
$A^p(L\ltimes J,D)$ consists of the antisymmetric maps $\omega:\Gamma(L)^p\rightarrow \Gamma(E)$ 
which are $C^{\infty}(M)$--multilinear in the sense \eqref{LAAH*4}.
The differential $d_{L\ltimes J,D}$ is then given again by \eqref{LAAH*1}
with $l_{\rho(s_i)}$ replaced by $D_{s_i}$. Similarly, in the supergeometric formulation,  
$d_{L\ltimes J,D}$  is given again by \eqref{LAAH*1'} with $l_{\rho(e_i)}$ replaced 
by $D_{e_i}$ and $f^k{}_{ij}$ by $f^k{}_{ij}\circ J$.

To an action of $L$ on $X$ along $J$, there is canonically associated a representation 
of the action Lie algebroid $L\ltimes J$ defined as follows.
Let $T^JX=\ker TJ$. Since $J$ is a submersion, $T^JX$ is a vector subbundle of $TX$.
Then, letting $E=T^JX$, $D_sv=l_{u(s)}v$, with $s\in \Gamma(L)$ and $v\in \Gamma(T^JX)$,
defines a representation $D_1$ of $L\ltimes J$. The restriction to $T^JX$ 
is required by the fulfillment of \eqref{LAAH*5}. In the same way, one can construct more general
representations $D_q$ of $L\ltimes J$ with $E=\wedge^qT^JX$. 
These are the only representations, which we shall consider in the following. 

Let $P$ be a Poisson structure on $X$ (cf. sect. \ref{sec:LAA}). 
As is well-known, $P$ is characterized its {\it Poisson cohomology}. 
This is the cohomology of the complex  $(V^*(X), d_P)$, where $V^q(X)=\Gamma(\wedge^q TX)$
is the space of $q$--vector fields and the Lichnerowicz differential 
$d_P:V^q(X)\rightarrow V^{q+1}(X)$ is defined by
\begin{equation}\label{LAAH*6}
d_PU=-[P,U]_{\wedge^*TX},
\end{equation}
$[\cdot,\cdot]_{\wedge^*TX}$ being the Schouten--Nijenhuis bracket. The cotangent bundle $T^*X$ 
of $X$ has a canonical Lie algebroid structure associated to the Poisson structure $P$ \cite{Vaisman}.
The Poisson cohomology of $X$ equals the Lie algebroid cohomology of $T^*X$. 

Let $J:X\rightarrow M,P$ be a fibered Poisson manifold (cf. sect. \ref{sec:LAA}). 
Then, by \eqref{LAA22}, $P\in \Gamma(\wedge^2T^JX)$. Define $V_J{}^q(X)=\Gamma(\wedge^q T^JX)$. Then,
$(V_J{}^*(X),d_P)$ is a subcomplex of the complex $(V^*(X), d_P)$ and, thus, itself a complex.
Its cohomology is the {\it tangential Poisson cohomology}. Here, the term ``tangential''
refers to the foliation of $X$ induced by $J$. 

Suppose that a Lie algebroid $L$ over $M$ acts infinitesimally on a fibered Poisson manifold 
$J:X\rightarrow M,P$ leaving $P$ invariant (cf. sect. \ref{sec:LAA}). 
Define $A_J{}^{p,q}(L)=\Gamma(\wedge^p(J^*L)^*\otimes\wedge^q T^JX)$.
For fixed $q$, $A_J{}^{p,q}(L)=A^p(L\ltimes J,D_q)$. Setting 
$d_{J,L}=d_{L\ltimes J,D_q}$, one has that $(A_J{}^{*,q}(L),d_{J,L})$ is a cochain complex.
For fixed $p$, if $\omega\in A_J{}^{p,q}(P)$, then $d_P\omega\in A_J{}^{p,q+1}(P)$, as is easy to verify using 
again \eqref{LAA22}. Thus, $(A_J{}^{p,*}(L),d_P)$ is cochain complex.
It can be verified that $d_Pd_{J,L}+d_{J,L}d_P=0$. It follows that $(A_J{}^{*,*}(L),d_{J,L},d_P)$ is 
a double cochain complex. 
We call the associated cohomology the {\it action Lie algebroid Poisson cohomology} of $L,J:X\rightarrow M,P$.
The cohomologies of $(A_J{}^{*,0}(L),d_{J,L})$,
$(A_J{}^{0,*}(L),d_P)$ are the action Lie algebroid cohomology of $L$ and the tangential Poisson cohomology 
of $P$, respectively. The total action Lie algebroid Poisson cohomology 
is the cohomology of the complex $(A_J{}^*(L),d_{J,L,P})$, where $A_J{}^*(L)$ is the complex $A_J{}^{*,*}(L),$ 
graded according to total degree, and $d_{J,L,P}=d_{J,L}+d_P$ is the total differential. 
See app. \ref{app:diff} for a supergeometric description of the double complex 
$(A_J{}^{*,*}(L),d_{J,L},d_P)$. 

\vfill\eject

\section{\normalsize \textcolor{blue}{BV cohomology of  the Lie algebroid 
Poisson sigma model}}\label{sec:LAcohom}

\par
\hskip.6cm The BV cohomology of the Lie algebroid Poisson sigma model is the cohomology of the nilpotent BV field 
variation operator $\delta_{J,P}$ (cf. eqs. \eqref{LAsigma6}). Since our sigma model is essentially 
a Poisson sigma model on the graded manifold $\mathfrak{X}_{L,J}$ (cf. eq. \eqref{LAsigma1}), one 
expects the BV cohomology to be related to the Poisson cohomology of the target space Poisson 
structure $\Pi$ (cf. eq. \eqref{LAsigma9}). One expects also there to be corrections 
due to the twisting by the moment map potential term. However, this point of view is not going to
yield much in the way of detailed cohomological information. Therefore, we shall not pursue it 
any longer.

To bring to focus the relation of the BV cohomology with the target space geometry of 
the sigma model, it is convenient to consider, instead of the BV variation operator $\delta_{J,P}$,
the mod $\mathbfs{d}$ BV variation operator
\begin{align}\label{LAcohom1}
\bar\delta_{J,P}=\delta_{J,P}-\mathbfs{d}.
\end{align}
As $\delta_{J,P}$, $\bar\delta_{J,P}$ is nilpotent \hphantom{xxxxxxxxxxxxxxxxxxxxxx}
\begin{align}\label{LAcohom2}
\bar\delta_{J,P}{}^2=0.
\end{align}
The cohomology of $\bar\delta_{J,P}$ is the mod $\mathbfs{d}$ BV cohomology and is the object 
of our study.

Because of the presence of the $0$-- and $-1$--degree superfields $\mathbfs{b}_i$ 
and $\mathbfs{B}_\alpha$, at each degree the most general superfield involves 
an infinite number of target space background fields. 
This renders the study of this cohomology problematic and not 
particularly illuminating. Fortunately, there is a subset $\matheul{X}^*$ of 
superfields that is interesting, on one hand, and is sufficiently restricted to allow 
for a simple study of the cohomology, on the other. $\matheul{X}^*$ consists
of the superfields of the form
\begin{align}\label{LAcohom3}
\mathbfs{\Phi}=\sum_{p,h,q}\frac{1}{p!h!q!}\Phi_{(p,h,q)i_1\dots i_p\alpha_1\ldots \alpha_h}{}^{a_1\ldots a_q}
(\mathbfs{x})\mathbfs{c}^{i_1}\ldots \mathbfs{c}^{i_p}
\mathbfs{C}^{\alpha_1}\ldots \mathbfs{C}^{\alpha_h}\mathbfs{y}_{a_1}\ldots\mathbfs{y}_{a_q},
\end{align}
where $\Phi_{(p,h,q)}\in \Gamma(\wedge^p(J^*L)^*\otimes \vee^h(J^*\ker\rho)^*\otimes\wedge^qT^JX)$.
Restricting to $T^JX$ (cf. sect. \ref{sec:LAH*}) amounts to the condition 
\begin{align}\label{LAcohom4}
\Phi_{(p,h,q)i_1\dots i_p\alpha_1\ldots \alpha_h}{}^{a_1\ldots a_{q-1}b}\partial_bJ^r=0.
\end{align}
\eqref{LAcohom4} is required by the target space global definedness
of the right hand side of \eqref{LAcohom3}, as follows easily from eqs. \eqref{chcoord*}. 
It also implies that $\matheul{X}^*$ is closed under the action of $\bar\delta_{J,P}$, 
as is apparent from eqs. \eqref{LAsigma6}. Thus, $\matheul{X}^*$ is a subcomplex 
of the mod $\mathbfs{d}$ BV cohomology superfield complex.

Using \eqref{LAsigma6}, one obtains straightforwardly the conditions on the $\Phi_{(p,h,q)}$ entailed 
by the mod $\mathbfs{d}$ BV cocycle condition $\bar\delta_{J,P}\mathbfs{\Phi}=0$.
The conditions are most naturally expressed by viewing the $\Phi_{(p,h,q)}$ 
as maps $\Phi_{(p,h,q)}:\Gamma(L)^p\times \Gamma(\ker\rho)^h
\rightarrow \Gamma{}_J(\wedge^qT^JX)$ antisymmetric in the first $p$ arguments and symmetric 
in the last $h$ arguments and $C^\infty(M)$--linear in the same sense as \eqref{LAAH*4}.
However, the resulting expressions 
are not very illuminating in the general case and, so, we shall not write them down explicitly.
Rather, we shall consider the first few low degree cases, because of their special interest. 

{\it Degree 0.} 

If $\mathbfs{\Phi}\in \matheul{X}^0$, then it is of the form
\begin{equation}\label{LAcohom7}
\mathbfs{\Phi}=\phi(\mathbfs{x}),
\end{equation}
where $\phi\in C^\infty(X)$. Imposing $\bar\delta_{J,P}\mathbfs{\Phi}=0$ leads to the 
equations
\begin{subequations}\label{LAcohom8}
\begin{align}
&\#_Pd\phi=0,
\vphantom{\bigg]}
\label{LAcohom8a}
\\
&l_{u(s)}\phi=0,
\vphantom{\bigg]}
\label{LAcohom8b}
\end{align}
\end{subequations}
with $s\in \Gamma(L)$. 

From a cohomological point of view, 
\eqref{LAcohom8} states that $\phi$ is a 0--cocycle of the total 
action Lie algebroid Poisson cohomology complex 
(cf. sect. \ref{sec:LAH*}). In more conventional terms, 
$\phi$ is a Casimir function of the Poisson structure $P$ 
invariant under the action of $L$.

In the extremal case $a$ of sect. \ref{sec:LAA}, \eqref{LAcohom8a}
is trivially satisfied as $P=0$ and \eqref{LAcohom8b} reduces into 
$l_{\rho(s)}\phi=0$. In the extremal case $b$, \eqref{LAcohom8}
states that $\phi$ is a Casimir function invariant under the action of the Lie algebra
associated to $L$.

{\it Degree 1.} 

If $\mathbfs{\Phi}\in \matheul{X}^1$, then it is of the form
\begin{equation}\label{LAcohom9}
\mathbfs{\Phi}=w^a(\mathbfs{x})\mathbfs{y}_a+\sigma_i(\mathbfs{x})\mathbfs{c}^i,
\end{equation}
where $w\in \Gamma(T^JX)$, $\sigma\in \Gamma((J^*L)^*)$. 
Imposing $\bar\delta_{J,P}\mathbfs{\Phi}=0$ leads to a set of equations, which can be cast as
\begin{subequations}\label{LAcohom10}
\begin{align}
&-[P,w]_{\wedge^*TX}=0,          
\vphantom{\bigg]}
\label{LAcohom10a}
\\
&l_{u(s)}w-\#_Pd\sigma(s)=0,
\vphantom{\bigg]}
\label{LAcohom10b}
\\
&l_{u(s)}\sigma(t)-l_{u(t)}\sigma(s)-\sigma([s,t])=0,
\vphantom{\bigg]}
\label{LAcohom10c}
\\
&l_w\mu(z)+\sigma(z)=0,
\vphantom{\bigg]}
\label{LAcohom10d}
\end{align}
\end{subequations}
with $s,t\in \Gamma(L)$ and $z\in \Gamma(\ker\rho)$. 

In cohomological terms, \eqref{LAcohom10a}--\eqref{LAcohom10c} state that
$(w,\sigma)$ is a $1$--cocycle of the total action Lie algebroid Poisson cohomology 
complex. In particular,  by \eqref{LAcohom10a}, $w$ is a 1--cocycle of the tangential Poisson 
cohomology complex of $P$ and,
by \eqref{LAcohom10c}, $\sigma$ is a $1$--cocycle of the action Lie algebroid 
cohomology complex (cf. sect. \ref{sec:LAH*}). \eqref{LAcohom10d} is a ``boundary condition''
determining $\sigma(z)$ for  $z\in \Gamma(\ker\rho)$. 
More conventionally, since $[w,P]_{\wedge^*TX}=l_wP$, \eqref{LAcohom10a} states that $w$ 
is a Poisson vector field of the Poisson structure $P$, i. e. a vector field whose flow leaves 
$P$ invariant. When $\sigma=0$, the flow leaves also invariant the moment map $\mu$
and the Lie algebroid action vector fields $u(s)$ for $s\in \Gamma(L)$.

In the extremal case $a$ of sect. \ref{sec:LAA}, one has not only that $P=0$ but also 
that $w=0$, by \eqref{LAcohom4}. 
Thus, \eqref{LAcohom10a}, \eqref{LAcohom10b} are trivially satisfied. 
Further, by \eqref{LAcohom10c}, being $u(s)=\rho(s)$, $\sigma$ is a 
1--cocycle of the Lie algebroid cohomology complex and, by \eqref{LAcohom10d}, 
the restriction of $\sigma$ to $\Gamma(\ker\rho)$ is trivial. 
In the extremal case $b$, \eqref{LAcohom10a} is the only non trivial condition.
Indeed, as $\ker\rho=L$, $\sigma(s)$ is expressed in terms of $w$ and $\mu(s)$ 
for all $s\in \Gamma(L)$ by \eqref{LAcohom10d} and \eqref{LAcohom10b}, \eqref{LAcohom10c} 
are automatically satisfied if \eqref{LAcohom10a} is.  

{\it Degree 2.} 

If $\mathbfs{\Phi}\in \matheul{X}^2$, then it is of the form
\begin{equation}\label{LAcohom12}
\mathbfs{\Phi}=\frac{1}{2}Q^{ab}(\mathbfs{x})\mathbfs{y}_a\mathbfs{y}_b
-v_i{}^a(\mathbfs{x})\mathbfs{c}^i\mathbfs{y}_a+\frac{1}{2}\tau_{ij}(\mathbfs{x})\mathbfs{c}^i\mathbfs{c}^j
+\nu_\alpha(\mathbfs{x})\mathbfs{C}^\alpha,
\end{equation}
in which  $Q\in \Gamma(\wedge^2T^JX)$, $v\in \Gamma((J^*L)^*\otimes T^JX)$,
$\tau\in \Gamma(\wedge^2(J^*L)^*)$ and $\nu\in \Gamma((J^*\ker\rho)^*)$.
Imposing $\bar\delta_{J,P}\mathbfs{\Phi}=0$ leads to the equations
\begin{subequations}\label{LAcohom14}
\begin{align}
&-[P,Q]_{\wedge^*TX}=0,
\vphantom{\bigg]}
\label{LAcohom14a}
\\
&l_{u(s)}Q-[P,v(s)]_{\wedge^*TX}=0,                
\vphantom{\bigg]}
\label{LAcohom14d}
\\
&-l_{u(s)}v(t)+l_{u(t)}v(s)+v([s,t])+\#_Pd\tau(s,t)=0,
\vphantom{\bigg]}
\label{LAcohom14c}
\\
&l_{u(r)}\tau(s,t)-l_{u(s)}\tau(r,t)+l_{u(t)}\tau(r,s)
\vphantom{\bigg]}
\label{LAcohom14b}
\\
&\hskip 4cm-\tau([r,s],t)+\tau([r,t],s)-\tau([s,t],r)=0,
\vphantom{\bigg]}
\nonumber
\\
&\#_Qd\mu(z)+\#_Pd\nu(z)-v(z)=0,
\vphantom{\bigg]}
\label{LAcohom14f}
\\
&l_{u(s)}\nu(z)+l_{v(s)}\mu(z)-\nu([s,z])-\tau(s,z)=0,
\vphantom{\bigg]}
\label{LAcohom14e}
\end{align}
\end{subequations}
where $r,s,t\in \Gamma(L)$ and  $z\in \Gamma(\ker\rho)$.

From a cohomological point of view, \eqref{LAcohom14a}--\eqref{LAcohom14b} state 
that $(Q,-v,\tau)$ is a 2--cocycle of the total action Lie algebroid Poisson cohomology.
In particular, by \eqref{LAcohom14a}, $Q$ is a $2$--cocycle of the
tangential Poisson cohomology complex of $P$ and, by \eqref{LAcohom14b},  
$\tau$ is a 2--cocycle of the action Lie algebroid cohomology complex. 
\eqref{LAcohom14f}, \eqref{LAcohom14e} are boundary conditions
determining $v(z)$ and $\tau(s,z)$ for $s\in \Gamma(L)$ and $z\in \Gamma(\ker\rho)$. 
In more conventional terms, setting $P'=P+Q$, $u'(s)=u(s)+v(s)$ and
$\mu'(z)=\mu(z)+\nu(z)$, \eqref{LAcohom14a}, \eqref{LAcohom14d}, \eqref{LAcohom14f}
state that $P'$ is a Poisson structure invariant under the flow of the vector fields
$u'(s)$ and that $\mu'$ is a (non equivariant) moment map for $u'$                    
to linear order in $Q$, $v$ and $\nu$ (cf. eqs. \eqref{LAA18}),\eqref{LAA24}, \eqref{LAA28}). 
When $\tau=0$, \eqref{LAcohom14c}, \eqref{LAcohom14e} state further that 
$u'$ defines a new action of $L$ on $X$ along $J$ and that the moment map
$\mu'$ is equivariant under the new action again to linear order in $Q$, $v$ and $\nu$ (cf. eqs. 
\eqref{LAA13}, \eqref{LAA27}). 

In the extremal case $a$ of sect. \ref{sec:LAA}, one has not only that $P=0$ but also 
that $Q=0$, $v=0$, by \eqref{LAcohom4}. Thus, \eqref{LAcohom14a}, \eqref{LAcohom14d}, \eqref{LAcohom14c},  
\eqref{LAcohom14f} are trivially satisfied.
Further, by \eqref{LAcohom14b}, being $u(s)=\rho(s)$, $\tau$ is a 2--cocycle 
of the Lie algebroid cohomology complex and, by \eqref{LAcohom14e}, 
the restriction of $\tau$ to $\Gamma(\ker\rho)$ is trivial. 
In the extremal case $b$,  \eqref{LAcohom14a} is the only non trivial condition.
Indeed, as $\ker\rho=L$, $v(s)$, $\tau(s,t)$ are expressed in terms of $Q$, $\nu(s)$ 
and $P$, $\mu(s)$ for all $s,t\in \Gamma(L)$ by \eqref{LAcohom14f}, \eqref{LAcohom14e}
and \eqref{LAcohom14d}--\eqref{LAcohom14b} are automatically satisfied 
if \eqref{LAcohom10a} is.  

The mod $\mathbfs{d}$ BV cohomology in higher degree is expected to exhibit 
a similar structure.
As usual, the degree 1 and 2 mod $\mathbfs{d}$ cohomologies relate to the infinitesimal symmetries
and infinitesimal deformations of the target space geometry, respectively. 
Strictly speaking this holds only when the 
action Lie algebroid 1-- and 2--cocycles $\sigma$ and $\tau$ above vanish.
The interpretation of $\sigma$ and $\tau$ in the general case  
is as yet unclear and calls for further investigation. 
 
It is interesting to compare the mod $\mathbfs{d}$ BV cohomology of the Lie algebroid 
Poisson sigma model in the case where $M$ is a point and $L$ is a Lie algebra
with that of the Poisson--Weil sigma model studied in ref. \cite{Zucchini7}. To begin with, 
one must recall that, in the situation considered, the Lie algebroid Poisson sigma model 
reproduces the Poisson--Weil sigma model for a trivial twisting principal bundle
(cf. sect. \ref{sec:LAsigma}). In the Poisson--Weil model, in the general case, the superfield 
$\mathbfs{c}^i$ is a generalized connection and must be absent in any expansion of the form
\eqref{LAcohom3} to have a superfield globally defined on the world-sheet.
Further, the coefficients of the expansion must be covariant under the action of the  
symmetry Lie group to have a superfield invariantly defined in target space.
For this reason, the analysis of ref. \cite{Zucchini7} was limited to the sector
of the mod $\mathbfs{d}$ BV cohomology complex formed by the superfields of the form 
\eqref{LAcohom3} with no $\mathbfs{c}^i$ factors and covariant coefficients
\footnote{$\vphantom{\bigg]}$
However, the analysis could have been generalized by introducing a fixed background generalized 
connection $\mathbfs{A}^i$ and replacing $\mathbfs{c}^i$ by $\mathbfs{c}^i-\mathbfs{A}^i$.}.
Therefore, the comparison of Lie algebroid Poisson sigma model and Poisson--Weil sigma model 
mod $\mathbfs{d}$ BV cohomologies can be carried out at best only upon restricting to 
a suitable sector of the former, that spanned by the superfields $\mathbfs{\Phi}$ 
of the form \eqref{LAcohom3} with no
$\mathbfs{c}^i$ occurrences. By \eqref{LAcohom9}, in degree $1$, this amounts to imposing that 
$\sigma=0$. Inspection of the 1--cocycle condition \eqref{LAcohom10} shows that 
$\mathbfs{\Phi}$ is a 1--cocycle of the equivariant Poisson cohomology (in the Cartan model),
as found in \cite{Zucchini7}. Similarly, by \eqref{LAcohom12}, in degree $2$,  one must have 
$v=0$ and $\tau=0$ and the 2--cocycle condition \eqref{LAcohom14} shows that 
$\mathbfs{\Phi}$ is a 2--cocycle of the equivariant Poisson cohomology, again as found
in \cite{Zucchini7}.

\vfill\eject

\section{\normalsize \textcolor{blue}{Hamiltonian Lie groupoid actions and Poisson reduction}}
\label{sec:LGG}

\par
\hskip.6cm The notion of Lie groupoid is related to that of Lie algebroid in 
the same way as the notion of Lie group is related to that of Lie algebra \cite{Mackenzie1,Cartier1}. 
Unlike what happens for Lie algebras and groups, not all Lie algebroids integrate 
to a Lie groupoid. In this section, we review briefly the theory of Lie groupoids and their associated 
Lie algebroids and of Hamiltonian actions of Lie groupoids on fibered Poisson manifolds.

A {\it groupoid} consists of two sets $G$ and $M$ and five maps 
$\alpha:G\rightarrow M$, $\beta:G\rightarrow M$, $1:M\rightarrow G$,
$\iota:G\rightarrow G$, $\mu:G\,{}_\alpha\!\!\times_\beta G\rightarrow G$, where 
$G\,{}_\alpha\!\!\times_\beta G=\{(g,h)\in G\times G|\alpha(g)=\beta(h)\}$ with the following 
properties. 

\par\noindent 
~~~1) For $m\in M$, $\alpha(1_m)=\beta(1_m)=m$.

\par\noindent 
~~~2) For $g\in G$, $\alpha(g^{-1})=\beta(g)$ and $\beta(g^{-1})=\alpha(g)$.

\par\noindent 
~~~3) For $(g,h)\in G\,{}_\alpha\!\!\times_\beta G$, $\alpha(gh)=\alpha(h)$ and $\beta(gh)=\beta(g)$.

\par\noindent 
~~~4) For $g\in G$, $g1_{\alpha(g)}=1_{\beta(g)}g=g$.

\par\noindent 
~~~5) For $g\in G$, 
$g^{-1}g=1_{\alpha(g)}$, $gg^{-1}=1_{\beta(g)}$.

\par\noindent 
~~~6) For $(g,h), (h,k)\in G\,{}_\alpha\!\!\times_\beta G$, $g(hk)=(gh)k$. 

\par\noindent 
Above, the standard notation $\iota(g)=g^{-1}$, $\mu(g,h)=gh$
is used. The structural maps $\alpha$, $\beta$, $1$, $\iota$, $\mu$ 
are called source, target, unit, 
inversion and partial multiplication, respectively. For $m\in M$, one sets
$G_m=\alpha^{-1}(m)$, $G^m=\beta^{-1}(m)$ and, for $m,n\in M$, $G_m^n=G_m\cap G^n$.

A groupoid $G$ over $M$ is a {\it Lie groupoid} if $G$ and $M$ are smooth manifolds
and the maps $\alpha$, $\beta$, $1$, $\iota$, $\mu$ are smooth with 
$\alpha$, $\beta$ surjective submersion,
$1$ and injective immersion and $\iota$ a diffeomorphism. In what follows, we shall consider 
exclusively Lie groupoids.

The prototype Lie groupoid over $M$ is the pair groupoid $G=M\times M$, whose  
structure maps are defined by 
$\alpha(m,n)=n$, $\beta(m,n)=m$, $1_m=(m,m)$, $(m,n)^{-1}=(n,m)$ and 
$(m,n)(n,p)=(m,p)$. 
Lie groupoids generalize Lie groups: a Lie group can be viewed as 
a Lie groupoid over the singleton manifold $M=\mathrm{pt}$.

A Lie groupoid $G$ is called {\it regular} if, for each $m\in M$, the target map $\beta$
restricts to a map $\beta: \alpha^{-1}(m)\rightarrow M$ of locally constant rank.
A Lie groupoid $G$ is called {\it transitive} if the map $(\alpha,\beta):G\rightarrow M\times M$
is a surjective submersion. 
Every transitive Lie groupoid is regular.

A {\it (base preserving) morphism} of two Lie groupoids $G$, $G'$ over $M$ is a smooth  
map $F:G\rightarrow G'$ such that 

\par\noindent 
~~~1) $\alpha'\circ F=\alpha$, $\beta'\circ F=\beta$ 

\par\noindent 
~~~3) For $(g,h)\in G\,{}_\alpha\!\!\times_\beta G$, $F(gh)=F(g)F(h)$. 

If $H$, $G$ are two Lie groupoids over $M$ and $H$ is an immersed submanifold 
of $G$, then $H$ is a {\it Lie subgroupoid} of $G$ if the natural injection
$I:H\rightarrow G$ is a Lie groupoid morphism.

Let $G$ be a regular Lie groupoid over $M$. Then, for $m\in M$, $G_m^m$ is a Lie group, 
the {\it isotropy group} of $m$. The {\it isotropy groupoid} of $G$, $I_G$, 
is defined as the union of all isotropy groups 
of $G$: \hphantom{xxxxxxxxxxxxxxxxxxxxxxxxxxx}
\begin{equation}\label{LGG6}
I_{G}=\hbox{$\bigcup$}_{m\in M}G_m^m. 
\end{equation}
With the structural maps and the differential structure inherited from $G$, 
$I_{G}$ is a Lie groupoid and a Lie subgroupoid of $G$.  $I_G$ is also a bundle of Lie groups.

Just as to any Lie group there is canonically associated a Lie algebra, to any Lie groupoid 
$G$ over $M$ there is canonically associated a Lie algebroid $AG$ over $M$. 
Explicitly, one has \hphantom{xxxxxxxxxxxxxxxxxxxxxxxx}
\begin{equation}\label{LGG1}
AG=\hbox{$\bigcup$}_{m\in M} T_{1_m}G_m
\end{equation}
with the vector bundle structure induced by that of $TG$. 
The Lie algebroid structure of $AG$ is defined as follows. A vector field $X\in \Gamma(TG)$ 
is said right invariant if: 
1) for $g\in G$, $X(g)\in T_gG_{\alpha(g)}$; 
2) for $(g,h)\in G{}_\alpha\!\!\times_\beta G$, $X(gh)=T_gR_h X(g)$, where, 
for $h\in G$, we define $R_h(g)=gh$ with $g\in G_{\beta(h)}$. One can show that there is a
one--to--one correspondence between sections of $AG$ and right
invariant vector fields of $G$ defined by
\begin{equation}\label{LGG2}
\tilde s(g)=T_{1_{\beta(g)}}R_g \,s(1_{\beta(g)}),\qquad g\in G,
\end{equation}
with $s\in \Gamma(AG)$. The Lie bracket of two right invariant vector fields of
$G$ is also right invariant. This allows to define the Lie bracket $[s,t]$ of two sections
$s,t \in \Gamma(AG)$ through the relation  
\begin{equation}\label{LGG3}
\widetilde {[s,t]}=[\tilde s, \tilde t].
\end{equation}
The anchor $\rho$ is defined by
\begin{equation}\label{LGG4}
\rho(s)(m)=T_{1_m}\beta \,s(1_m),
\end{equation}
for $m\in M$ and $s\in \Gamma(AG)$. It is straightforward to check that the basic
relations \eqref{LAA1}--\eqref{LAA3} are satisfied.

For the pair groupoid $G=M\times M$, $AG=TM$. If $G$ is a Lie group, then $AG=\mathfrak{g}$,
the usual Lie algebra of $G$. 

If $G$ is a regular Lie groupoid, then $AG$ is a regular Lie algebroid. Similarly, if $G$ is a 
transitive Lie groupoid, then $AG$ is a transitive Lie algebroid (cf. sect. \ref{sec:LAA}). 

Let $G$, $G'$ be two Lie groupoids over $M$ and let $F:G\rightarrow G'$ be a groupoid morphism. 
Then, setting \hphantom{xxxxxxxxxxxxxxxxxxxxxxxx}
\begin{equation}
F_*(s)_m=T_{1_m}F\,s_m,
\end{equation}
with $s_m\in T_{1_m}G_m$, defines a Lie algebroid morphism (cf. eq. \eqref{LAAmorph1}, 
\eqref{LAAmorph2}). 

If $H$ is a Lie subgroupoid of $G$, then $AH$ is a Lie subalgebroid of $AG$ (cf. sect. \ref{sec:LAA}). 
In particular, if $G$ is a regular Lie groupoid, $AI_G$ is a Lie subalgebroid of
$AG$. In fact, one has $AI_G=\ker\rho$, as follows easily from \eqref{LGG4}.

Let $G$ be a Lie groupoid over $M$ and let $J:X\rightarrow M$ be a fibered manifold.
A {\it left action} of $G$ on $X$ along $J$ is a smooth 
map $\lambda:G\,{}_\alpha\!\!\times_J X\rightarrow X$, where
$G\,{}_\alpha\!\!\times_J X=\{(g,x)\in G\times X|\alpha(g)=J(x)\}$ with the following 
properties. 

\par\noindent 
~~~1) For $(g,x)\in G\,{}_\alpha\!\!\times_J X$, $J(gx)=\beta(g)$.

\par\noindent 
~~~2) For $x\in X$, $1_{J(x)}x=x$.

\par\noindent 
~~~3) For $(g,h)\in  G\,{}_\alpha\!\!\times_\beta G$, $(h,x)\in G\,{}_\alpha\!\!\times_J X$,
$g(hx)=(gh)x$.

\par\noindent 
Above, the standard notation $\lambda(g,x)=gx$ is used. 

To a left action of $G$ on $X$ along $J$, there is 
canonically associated an infinitesimal action of $AG$ on 
$X$ along $J$ (cf. sect. \ref{sec:LAA}). The associated map $u:\Gamma(AG)\rightarrow \Gamma(TX)$ 
is defined by
\begin{equation}\label{LGG5}
u(s)(x)=T_{1_{J(x)}}A_x \,s(1_{J(x)}),\qquad x\in G,
\end{equation}
for $s\in \Gamma(AG)$, where, for $x\in X$, we define $A_x(g)=gx$ with $g\in G_{J(x)}$.
It is a simple matter to check that the basic properties \eqref{LAA12}--\eqref{LAA14} of an 
infinitesimal action hold. 

To a left action of $G$ on $X$ along $J$, there is canonically associated a Lie groupoid 
structure over $X$ on the pull back $G\,{}_\alpha\!\!\times_J X$. The structural maps are defined by
$s((g,x))=x$, $t((g,x))=gx$ with $(g,x)\in  G\,{}_\alpha\!\!\times_J X$,
$1_x=(1_{J(x)},x)$ with $x\in X$, $(g,x)^{-1}=(g^{-1},gx)$  with $(g,x)\in  G\,{}_\alpha\!\!\times_JX$ and 
and $(g,x)(h,y)=(gh,y)$  with $(g,x),(h,y)\in  G\,{}_\alpha\!\!\times_J X$ such that 
$x=hy$.
The resulting Lie groupoid is called the {\it action Lie groupoid}
corresponding to the left action and is usually denoted by 
$G\ltimes J$.

It is an important result that $A(G\ltimes J)\simeq AG\ltimes J$:
the Lie algebroid of the action Lie groupoid $G\ltimes J$ is isomorphic to
the action Lie algebroid $AG\ltimes J$ (cf. sect. \ref{sec:LAA}, eqs.
\eqref{LAAaLa1}, \eqref{LAAaLa2}). 

Next, we discuss a generalization of Hamiltonian symmetry reduction for Lie groupoid 
actions on fibered Poisson manifolds. We follow closely the treatment of Bos in \cite{Bos1}
\footnote{$\vphantom{\bigg[}$ Actually, Bos considers only the symplectic case. 
Further, he uses the more precise term
strongly internally Hamiltonian in place of Hamiltonian.}.

Let $G$ be a regular Lie groupoid over $M$ acting on a fibered Poisson manifold
$J:X\rightarrow M,P$ (cf. eqs. \eqref{LAA18}, \eqref{LAA22}).
$P$ is said {\it invariant}, if it is invariant under the associated 
infinitesimal action of $AG$ (cf. eq. \eqref{LAA24}).
For $P$ invariant, the action is said Hamiltonian, if there exists an 
equivariant moment map $\mu$ for the $AG$ action (cf. eqs. \eqref{LAA26}--\eqref{LAA28}). 
Henceforth, we assume that $P$ is invariant and that the action is Hamiltonian
with moment map $\mu$.

As $I_G$ is a Lie subgroupoid of $G$ and $AI_G=\ker\rho$, one can
view the moment map as a map $\mu:X\rightarrow (AI_G)^*$ such that $\pi_{AI_G}\circ \mu=J$,
where $\pi_{AI_G}:(AI_G)^*\rightarrow M$ is the bundle projection. Let $0$ denote the zero section of $(AI_G)^*$. 
Suppose that $0(M)\subset\im\mu$.
Suppose further that, for each $m\in M$, the Lie group $G_{m}^{m}$ is connected and 
that it acts freely and properly on $\mu^{-1}(0(m))$. 
Then, for each $m\in M$, the quotient manifold 
\begin{equation}
X_{m}:=G_{m}^{m}\backslash\mu^{-1}(0(m))
\end{equation} 
is a smooth manifold. 
Now, note that $\mu^{-1}(0(m))\subset J^{-1}(m)$. Then, since $P$ is fibered, $P\big|_{J^{-1}(m)}$ 
is a Poisson structure on $J^{-1}(m)$ (cf. eq. \eqref{LAA22}). Likewise, since the $u$ is projectable, 
$u(z)\big|_{J^{-1}(m)}$ is a vector field on $J^{-1}(m)$, for all $z\in \Gamma(\ker\rho)$
(cf. eq. \eqref{LAA14}). By the classic result of Marsden and Ratiu \cite{Ratiu1},
$X_{m}$ inherits a Poisson structure (Marsden-Weinstein quotient).

Suppose that $\mu$ and $0$ are transversal, i.e. that, for any $m\in M$ and
any $x\in X$ such that $\mu(x)=0(m)$, $T_m0(T_mM)$ and $T_x\mu(T_xX)$ are transversal
in $T_{0(m)}(AI_G)^*$.
Then, $\mu^{-1}(0(M))$ is a manifold. The map
\begin{equation}
\hbox{$\bigcup$}_{m\in M}X_{m}=I_{G}\backslash\mu^{-1}(0(M))\rightarrow M
\end{equation}
is a smooth family of Poisson manifolds. 

Thus, under the assumption listed above,
the Hamiltonian action of a regular Lie groupoid $G$ over $M$ 
on a Poisson manifold $X$ fibered over $M$ along the fibration leaving the 
Poisson structure invariant induces a family version of the Hamiltonian symmetry reduction. 
This reduction is encoded in the target space geometry of the Lie algebroid
Poisson sigma model described in sect. \ref{sec:LAsigma}, when the background Lie algebroid 
$L$ is the Lie algebroid $AG$ of the Lie groupoid $G$.

\vfill\eject

\section{\normalsize \textcolor{blue}{Examples}}\label{sec:exa}

In this section, we shall illustrate a class of examples of the target space geometry of the Lie
algebroid Poisson sigma model. The geometrical data are listed 
at the beginning of sect. \ref{sec:LAsigma}. 

Suppose that $X$ is a vector bundle over $M$.
There exists a vector bundle $\mathfrak{gl}(X)$ over $M$
that fits in a short exact sequence of base preserving vector bundle morphisms of the form
\hphantom{xxxxxxxxxxxxxxxxxx}
\begin{equation}\label{exactglX}
\xymatrix{0\ar[r]&\End X\ar[r]^\iota&\mathfrak{gl}(X)\ar[r]^\varpi&TM\ar[r]&0.}
\end{equation}
$\mathfrak{gl}(X)$ is isomorphic to the direct sum bundle $TM\oplus\End X$. The isomorphism is 
not canonical depending on the choice of a splitting, a vector bundle morphism 
$\sigma:TM\rightarrow \mathfrak{gl}(X)$ such that $\varpi\circ \sigma=\id_{TM}$.
The splittings are in one to one correspondence with the connections of $X$.
See \cite{Mackenzie1} for background.

The details of the local description of the vector bundle $\mathfrak{gl}(X)$ are
provided in app. \ref{app:covarianceX}. We denote by $m^r$ and $(\mu^r,\alpha^A{}_B)$ 
the local base and fiber coordinates of $\mathfrak{gl}(X)$. 
The morphisms $\iota$ and $\varpi$ are then given locally by 
$\iota(m,\alpha)=(m^r,0,\alpha^A{}_B)$ and 
$\varpi(m,\mu,\alpha)=(m^r,\mu^r)$, respectively.

$\mathfrak{gl}(X)$ has a natural structure of Lie algebroid. 
Its anchor is the morphism $\varpi$ appearing in the sequence \eqref{exactglX}.
Its Lie bracket is defined as follows.
Let $s,t\in \Gamma(\mathfrak{gl}(X))$ be locally given as 
$s(m)\simeq(v^r(m),s^A{}_B(m))$, $t(m)\simeq(w^r(m),t^A{}_B(m))$, 
respectively. Then,
\begin{align}
[s,t]&(m)
\label{[s,t]glX}
\\
&\simeq ((v^s\partial_sw^r-w^s\partial_sv^r)(m), (v^s\partial_st^A{}_B-w^s\partial_ss^A{}_B
-s^A{}_Ct^C{}_B+t^A{}_Cs^C{}_B)(m)).
\nonumber
\end{align}
The Lie algebroid $\mathfrak{gl}(X)$ is transitive and thus regular.

To any section $s\in \Gamma(\mathfrak{gl}(X))$, there is associated a 
{\it linear vector field} $u(s)$ on $X$ \cite{Mackenzie1}, that is a vector field depending 
linearly on the fiber coordinates of $X$, as follows.
Let $(m^r,e^A)$ and $(\mu^r,\epsilon^A)$ be local base and fiber coordinates of $TX$,
$m^r$ and $e^A$ being base and fiber coordinates of $X$
(cf. app. \ref{app:covarianceX}). If $s(m)\simeq(v^r(m),s^A{}_B(m))$ locally, then 
\begin{equation}\label{linsglX}
u(s)(m,e)\simeq(v^r(m),s^A{}_B(m)e^B).
\end{equation} 
The linear vector fields form a Lie subalgebra $\Gamma(TX)_{\mathrm{lin}}$ of
$\Gamma(TX)$ and the map $s\rightarrow u(s)$ defines a Lie algebra isomorphism
$\Gamma(\mathfrak{gl}(X))\simeq \Gamma(TX)_{\mathrm{lin}}$. Using this identification, 
one can view $\mathfrak{gl}(X)$ as a Lie algebroid whose sections are the linear
vector fields of $X$. 

To any section $s\in \Gamma(\mathfrak{gl}(X))$, there is associated a 
{\it derivative endomorphism} $\mathcal{D}_s$ of $X$. A derivative endomorphism is
an $\mathbb{R}$--linear map $\mathcal{D}:\Gamma(X)\rightarrow \Gamma(X)$ such that
there is a vector field $u_{\mathcal{D}}\in \Gamma(TM)$ such that
\begin{equation}\label{linendglX}
\mathcal{D}(f\sigma)=f\mathcal{D}\sigma+(l_{u_{\mathcal{D}}}f)\sigma
\end{equation} 
for $f\in C^\infty(M)$ and $\sigma\in \Gamma(X)$ \cite{Mackenzie1}. 
If $s(m)\simeq(v^r(m),s^A{}_B(m))$ locally, then 
\begin{equation}\label{linendglXexp}
\mathcal{D}_s\sigma^A(m)=(v^r\partial_r\sigma^A-s^A{}_B\sigma^B)(m).
\end{equation} 
Note that $u_{\mathcal{D}_s}=v$.
The derivative endomorphism form a Lie algebra $\mathfrak{D}_X$ if the Lie
bracket is defined as the usual operator commutator. The map $s\rightarrow \mathcal{D}_s$ 
defines a Lie algebra isomorphism $\Gamma(\mathfrak{gl}(X))\simeq \mathfrak{D}_X$.
Using this identification, one can view $\mathfrak{gl}(X)$ as a Lie algebroid whose sections 
are the derivative endomorphisms of $X$. Upon doing so, a representation of a Lie algebroid 
$L$ over $M$ on $X$ can be regarded 
as a Lie algebroid morphism $D:L\rightarrow \mathfrak{gl}(X)$
(cf. sect. \ref{sec:LAH*}, eq. \eqref{LAAH*2}). 

As $X$ is a vector bundle over $M$, $J:X \rightarrow M$ is a fibered manifold, 
$J$ being the bundle projection. In local coordinates, $J^r(m,e)=m^r$.

If $L$ is a Lie subalgebroid of $\mathfrak{gl}(X)$, then the map
$s\in \Gamma(L)\rightarrow u(s)\in \Gamma(TX)$ defines an infinitesimal action of
$L$ on $X$ along $J$ (cf. sect. \ref{sec:LAA}). Indeed, \eqref{LAA12}--\eqref{LAA14} 
are satisfied, as is easy to check from \eqref{[s,t]glX}, \eqref{linsglX}.   

If $X$ is fibered Poisson manifold and $P\in \Gamma(\wedge^2TX)$
is its Poisson structure, then \eqref{LAA23} holds.
In local coordinates, this yields the equations
\begin{equation}\label{Pra=0X}
P^{rs}=0, \qquad P^{rA}=0.
\end{equation} 
Only the components $P^{AB}$ may be non zero. Thus, actually 
$P\in \Gamma(\wedge^2 \Vrtc TX)$, where $\Vrtc TX=J^*X$ is the vertical subbundle of $TX$.
The Poisson condition \eqref{LAA21} obeyed by 
$P$ reduces then into 
\begin{equation}\label{LAA21X}
P^{AD}\partial_DP^{BC}+P^{BD}\partial_DP^{CA}+P^{CD}\partial_DP^{AB}=0.
\end{equation}
If $\partial_CP^{AB}=0$, \eqref{LAA21X} is automatically satisfied.
In that case, one can view $P\in \Gamma(\wedge^2 X)$.

Let $s\in \Gamma(\mathfrak{gl}(X))$ and let $P$ be invariant under the linear vector field
$u(s)$. 
Then, \eqref{LAA25} holds. Explicitly, if $s\simeq(v^r,s^A{}_B)$, one has 
\begin{equation}\label{PsinvX}
v^r\partial_rP^{AB}-s^A{}_CP^{CB}-s^B{}_CP^{AC}+s^C{}_De^D\partial_CP^{AB}=0.
\end{equation}
If $\varpi(s)=0$, then $u(s)$ is Hamiltonian if 
\begin{subequations}\label{HamXu(s)}
\begin{align}
&v^r=0
\label{HamXu(s)a}
\\
&s^A{}_Be^B=-P^{AB}\partial_B\mu(s),
\label{HamXu(s)b}
\end{align}
\end{subequations}
for some function $\mu(s)\in C^\infty(X)$.

In general, the sections $s\in \Gamma(\mathfrak{gl}(X))$ such that 
\eqref{PsinvX} holds are not sections of some regular Lie subalgebroid $L$ of 
$\mathfrak{gl}(X)$. If such an $L$ can be found, however, then the infinitesimal 
action of $L$ on $X$ leaves $P$ invariant (cf. sect. \ref{sec:LAA}, eq. \eqref{LAA24}). 
Even when $L$ does exist, in general $u(s)$ is not Hamiltonian 
for $s\in \Gamma(\ker\rho)$, where $\rho=\varpi\big|_L$ is the anchor of $L$.
We do not know any general condition ensuring the existence of $L$ and the 
Hamiltonianity of its action on $X$. Below, we present a possible scenario
where this can happen.

Suppose that $P\in \Gamma(\wedge^2TX)$ satisfies \eqref{Pra=0X}
and the linearity condition
\begin{equation}\label{PlinX}
P^{AB}(m,e)=\pi^{AB}{}_C(m)e^C,
\end{equation}
where $\pi\in\Gamma(\wedge^2X\otimes X^*)$. The Poisson condition \eqref{LAA21X}
then becomes a purely algebraic relation
\begin{equation}\label{LAA21piX}
\pi^{AD}{}_E\,\pi^{BC}{}_D+\pi^{BD}{}_E\,\pi^{CA}{}_D+\pi^{CD}{}_E\,\pi^{AB}{}_D=0.
\end{equation}
Thus, the dual bundle $X^*$ of $X$ is a bundle of Lie algebras.

If $s\in \Gamma(\mathfrak{gl}(X))$ with $s\simeq(v^r,s^A{}_B)$ locally,
\eqref{PsinvX} is satisfied if and only if 
\begin{equation}\label{pisinvX}
v^r\partial_r\pi^{AB}{}_C-s^A{}_D\pi^{DB}{}_C-s^B{}_D\pi^{AD}{}_C+s^D{}_C\pi^{AB}{}_D=0.
\end{equation}
This condition is purely algebraic in $s$. Thus, it defines a subspace
in each fiber of $\mathfrak{gl}(X)$. With some regularity assumption on $\pi$ made,
this distribution of subspaces is a subbundle $L$ of $\mathfrak{gl}(X)$.
Since $l_{u(s)}P=l_{u(t)}P=0$ implies $l_{u([s,t])}P=0$, $L$ is in fact 
a Lie subalgebroid of $\mathfrak{gl}(X)$. $L$ then acts infinitesimally on $X$ along $J$
leaving $P$ invariant, by construction.

If $s\in \Gamma(\ker\rho)$, then $v^r=0$ and \eqref{pisinvX} 
becomes 
\begin{equation}\label{pisinvXanch}
s^A{}_D\pi^{DB}{}_C+s^B{}_D\pi^{AD}{}_C-s^D{}_C\pi^{AB}{}_D=0.
\end{equation}
Then, for every $m\in M$, $s(m)$ is a 1--cocycle of the Lie algebra cohomology of $X^*{}_m$ with values in 
$\ad X^*{}_m$. If
\begin{equation}\label{strivial}
s^A{}_B=-\pi^{AC}{}_Bt_C
\end{equation}
for some $t\in \Gamma(X^*)$, this 1--cocycle is a 1--coboundary.
In such case, $u(s)$ is Hamiltonian: \eqref{HamXu(s)b} is fulfilled
with
\begin{equation}\label{mu(s)expr}
\mu(s)(m,e)=t_A(m)e^A.
\end{equation}
In order this to be the case, it suffices to require that the 
1st Lie algebra cohomology of $X^*{}_m$ vanishes for all $m\in M$.

The geometrical setup described above is automatically integrable. 
Let $GL(X)$ be the set of all linear isomorphisms $T:X_m\rightarrow X_n$
with $m,n\in M$. Then, $GL(X)$ has a natural structure of Lie groupoid over $M$:
$\alpha(T)=m$,  $\beta(T)=m$, for $T:X_m\rightarrow X_n$; $1_m=\id_{X_m}$;
the inversion and partial multiplication are the corresponding operations for linear 
isomorphisms. It can be shown that $AGL(X)\simeq \mathfrak{gl}(X)$
\cite{Mackenzie1,Mackenzie2}.
Thus, $\mathfrak{gl}(X)$ is automatically integrable and so is every Lie subalgebroid 
$L$ of $\mathfrak{gl}(X)$.

\vfill\eject

\section{\normalsize \textcolor{blue}{Concluding remarks}}
\label{sec:conclusions}

In this final section, we briefly review and comment on the problems which 
are still open.

When $M=\mathrm{pt}$ and $J:X\rightarrow M$ is the constant map, the Lie
algebroid $L$ is an ordinary Lie algebra acting infinitesimally on $X$ leaving
the Poisson structure $P$ invariant. As noticed in sect. \ref{sec:LAsigma}, in this case 
the Lie algebroid Poisson sigma model reduces into the Poisson--Weil model of 
refs.\cite{Zucchini6,Zucchini7} for trivial twisting 
bundle. 
The twisting bundle is a principal bundle on the world sheet $\Sigma$ with structure group 
$G$ integrating the Lie algebra $L$. It is the gauge bundle of the Poisson--Weil sigma model as 
a 2--dimensional gauge theory. 
The natural question arises whether it is possible to generalize
the construction described in the present work in such a way
to recover, in the Lie algebra case, the Poisson--Weil model 
with arbitrary twisting bundle. Presumably, this requires the following.

\begin{enumerate}

\item The symmetry of the target space geometry is encoded in a Lie 
groupoid $G$ over $M$ integrating $L$.

\item The twisting structure is a principal groupoid bundle $P$ with base 
$\Sigma$ and structure groupoid $G$.

\end{enumerate}
\noindent
Recall that  a {\it principal groupoid bundle} $P$ over a manifold $\Sigma$ 
is a smooth fiber bundle $\pi:P\rightarrow\Sigma$ endowed with a 
smooth right action $\mu$ of $G$ along $\kappa:P\rightarrow M$ preserving the fibers of $P$ and 
such that the map
\begin{equation}
(\pro_1,\mu): P{_{\kappa}\times_{\beta}}G \to 
P{_{\pi}\times_{\pi}} P : (p,g) \mapsto (p, pg) \; 
\end{equation}
is a diffeomorphism. Groupoid right actions are defined in a way totally analogous to left actions
(cf. sect. \ref{sec:LGG}).
Diagrammatically, the bundle can be represented as
\begin{equation*} 
\xymatrix{P\ar[dr]_{\kappa}\ar[d]_{\pi} 
&G\ar@<2pt>[d]^\beta\ar@<-2pt>[d]_\alpha\\ 
\Sigma & M}. 
\end{equation*}
It can be shown that when $M=\mathrm{pt}$ and $G$ is a Lie group, one recovers 
the customary notion of principal (group) bundle. See \cite{Stienon1}
and references therein for background.

Unfortunately, at present, it is not clear to us how to implement this more general 
form of twisting in a Lagrangian field theoretic framework essentially because
we do not know how to build objects globally defined on $\Sigma$ which can be integrated
out of the above geometrical data. This is an open issue calling for further 
investigation.

As is well known, the BV master action of a field theory is not directly usable
for quantization: gauge fixing is required. Fixing the gauge consists in restricting to a 
suitable Lagrangian submanifold if field space. It is notoriously a very difficult problem. 
Normally, it can be done only in certain cases, when the background geometry 
has extra structures, and there are no general methods for its implementation.

For the Poisson-Weil model, gauge fixing has been worked out by us in \cite{Zucchini7}, 
taking inspiration from the classical work of AKSZ \cite{AKSZ}, and has led 
to interesting topological field theories such as the 2--dimensional Donaldson--Witten 
topological gauge theory \cite{Witten5,Witten6} and the gauged A topological sigma model
\cite{Baptista1,Baptista2,Baptista3}.  
At the moment, we know no sensible gauge fixing prescriptions of the Lie algebroid Poisson 
sigma model yielding interesting topological field theories. 
As far as we know, there may not be any.

Generalized complex geometry \cite{Hitchin1,Gualtieri} has been the object of much interest 
in recent years for its role in superstring flux compactifications \cite{Grana}. 
In \cite{Zucchini1,Zucchini2,Zucchini3}, following the AKSZ philosophy of \cite{Cattaneo1,Cattaneo2}
and extending the Poisson sigma model, we introduced a BV field theoretic realization of generalized 
complex geometry, the Hitchin sigma model, and in \cite{Zucchini6}, we gauged it by coupling it to 
the Weil model. It would be interesting to generalize the construction of the present paper 
to the Hitchin model. The target space geometry of the ``Lie algebroid Hitchin model''
is expected to be extremely rich and interesting.


\vfill\eject

\appendix

\section{\normalsize \textcolor{blue}{Analysis of  covariance I}}
\label{app:covariance}

In this appendix, we shall present an analysis of covariance for the cotangent bundle
$T[1]^*\mathfrak{X}_{L,J}$ of the manifold $\mathfrak{X}_{L,J}$ defined in \eqref{LAsigma1}.

Recall that $L$ is a regular Lie algebroid over $M$ 
acting infinitesimally on a fibered manifold $J:X\rightarrow M$.
Since $\ker\rho$ is a subbundle of $L$, we conveniently use trivializations of $L$ 
adapted to $\ker\rho$. Thus, the fiber coordinates $\{v^i\}$ of any vector $v\in L$ 
get subdivided as $\{v^\alpha\}\cup\{v^\kappa\}$ and $v\in\ker\rho$
if and only if $v^\kappa=0$ for all $\kappa$. We denote by 
$(T^{i'}{}_j)$ the transition matrix function of a generic change of adapted trivialization
of $L$. The upper left block of $(T^{i'}{}_j)$, $(T^{\alpha'}{}_\beta)$, is the  
transition matrix function of the associated  change of trivialization of $\ker\rho$.

We denote by $\xi^a$, $(\beta_i,\Beta_\alpha)$ respectively 
the base and fiber coordinates of 
the vector bundle $\mathfrak{X}_{L,J}=(J^*L)^*[0]\oplus(J^*\ker\rho)^*[-1]$
with respect to some trivialization.
Then, the cotangent bundle $T^*[1]\mathfrak{X}_{L,J}$ has base coordinates 
$(\xi^a,\beta_i,\Beta_\alpha)$ and fiber coordinates $(\eta_a,\gamma^i,\Gamma^\alpha)$.

A straightforward differential geometric analysis shows that 
under a change of trivialization, one has
\begin{subequations}\label{chcoord*}
\begin{align}
&\xi^{a'}=F^{a'}(\xi),
\vphantom{\bigg[}
\label{chcoord*a}
\\
&\beta_{i'}=T^{-1j}{}_{i'}(J'(F(\xi)))\beta_j,
\vphantom{\bigg[}
\label{chcoord*b}
\\
&\Beta_{\alpha'}=T^{-1\beta}{}_{\alpha'}(J'(F(\xi)))\Beta_\beta.
\vphantom{\bigg[}
\label{chcoord*c}
\\
&\eta_{a'}=\partial_{a'}F^{-1b}(F(\xi))\big[\eta_b
+T^{-1i}{}_{k'}(J'(F(\xi)))\partial_bJ^r(\xi)\partial_rT^{k'}{}_j(J(\xi))\beta_i\gamma^j
\vphantom{\bigg[}
\label{chcoord*d}
\\
&\hskip4.1cm
+T^{-1\alpha}{}_{\gamma'}(J'(F(\xi)))\partial_bJ^r(\xi)\partial_rT^{\gamma'}{}_\beta(J(\xi))\Beta_\alpha\Gamma^\beta
\big],
\vphantom{\bigg[}
\nonumber
\\
&\gamma^{i'}=T^{i'}{}_j(J(\xi))\gamma^j,
\vphantom{\bigg[}
\label{chcoord*e}
\\
&\Gamma^{\alpha'}=T^{\alpha'}{}_\beta(J(\xi))\Gamma^\beta,
\vphantom{\bigg[}
\label{chcoord*f}
\end{align}
\end{subequations}

Under a change of trivialization, the anchor and structure functions
transform as follows
\begin{align}
u_{i'}{}^{a'}(\xi')&=\partial_bF^{a'}(\xi)T^{-1j}{}_{i'}(J'(F(\xi)))u_j{}^b(\xi),
\vphantom{\bigg[}
\label{}
\\
f^{k'}{}_{i'j'}(J'(\xi'))
&=T^{k'}{}_n(J(\xi))T^{-1l}{}_{i'}(J'(F(\xi)))T^{-1m}{}_{j'}(J'(F(\xi)))
\vphantom{\bigg[}
\label{}
\\
&\hphantom{=}\times\big[f^n{}_{lm}(J(\xi))
-T^{-1n}{}_{h'}(J'(F(\xi)))u_l{}^a(\xi)\partial_aJ^r(\xi)\partial_rT^{h'}{}_m(J(\xi))
\vphantom{\bigg[}
\nonumber
\\
&\hphantom{=}\hskip2.7cm
+T^{-1n}{}_{h'}(J'(F(\xi)))u_m{}^a(\xi)\partial_aJ^r(\xi)\partial_rT^{h'}{}_l(J(\xi))
\big].
\vphantom{\bigg[}
\nonumber
\end{align}

Exploiting the above relations, it is straightforward though lengthy to verify the target space global 
definedness of the sigma model action \eqref{LAsigma4}.

\vfill\eject

\section{\normalsize \textcolor{blue}{Analysis of  covariance II}}
\label{app:covarianceX}

In this appendix, we shall present an analysis of covariance for the vector bundle 
$\mathfrak{gl}(X)$ studied in sect. \ref{sec:exa}.

Let $X$ be a vector bundle over $M$. Let $m^r$, $e^A$ be the base and fiber coordinates
associated with a given trivialization of $X$, respectively. Under a change of trivialization, 
they transform as 
\begin{subequations}\label{chcoordX}
\begin{align}
&m^{r'}=\Phi^{r'}(m),
\vphantom{\bigg[}
\label{chcoordXa}
\\
&e^{A'}=\Theta^{A'}{}_B(m)e^B,
\vphantom{\bigg[}
\label{chcoordXb}
\end{align}
\end{subequations}
where $(\Theta^{A'}{}_B)$ is the transition matrix function of the trivialization change.

Consider next the vector bundle $TX$. To each trivialization of $X$, there corresponds one of $TX$, with  
associated base and fiber coordinates $(m^r,e^A)$, $(\mu^r,\epsilon^A)$, respectively. 
Under a change of trivialization, one has \eqref{chcoordX} and 
\begin{subequations}\label{chcoordTX}
\begin{align}
&\mu^{r'}=\partial_s\Phi^{r'}(m)\mu^s,
\vphantom{\bigg[}
\label{chcoordTXa}
\\
&\epsilon^{A'}=\partial_r\Theta^{A'}{}_B(m)\mu^re^B+\Theta^{A'}{}_B(m)\epsilon^B.
\vphantom{\bigg[}
\label{chcoordTXb}
\end{align}
\end{subequations}
$TX$ is actually a ``double vector bundle'' \cite{Mackenzie1}: it is not only a vector bundle over $X$
but also one over $TM$. This becomes apparent upon considering $(m^r,\mu^r)$ as base coordinates and
$(e^A,\epsilon^A)$ as fiber coordinates.

The vector bundle $\mathfrak{gl}(X)$ can be described locally by
specifying an atlas of local coordinates together with the coordinate
change transformation relations.
To each trivialization of $X$, there corresponds one of $\mathfrak{gl}(X)$
with base coordinates $m^r$ and fiber coordinates $\mu^r$, $\alpha^A{}_B$, transforming according to
\eqref{chcoordXa}, \eqref{chcoordTXa} and
\begin{equation}\label{chcoordglX}
\alpha^{A'}{}_{B'}=\partial_r\Theta^{A'}{}_C(m)\Theta^{-1C}{}_{B'}(\Phi(m))\mu^r
+\Theta^{A'}{}_C(m)\alpha^C{}_D\Theta^{-1D}{}_{B'}(\Phi(m)).
\end{equation}

A connection of $X$ is given in a trivialization by local $1$--forms $A_r{}^A{}_B(m)dm^r$ 
transforming under the  change of trivialization \eqref{chcoordX} as
\begin{align}
A_{r'}{}^{A'}{}_{B'}(m')=\partial_{r'}\Phi^{-1s}(\Phi(m))\big[-&\partial_s\Theta^{A'}{}_C(m)\Theta^{-1C}{}_{B'}(\Phi(m))
\vphantom{\bigg[}
\label{chcoordconnX}
\\
&+\Theta^{A'}{}_C(m)A_s{}^C{}_D(m)\Theta^{-1D}{}_{B'}(\Phi(m))\big].
\vphantom{\bigg[}
\nonumber
\end{align}
Upon picking a connection, there is defined a vector bundle isomorphism
$\mathfrak{gl}(X)\simeq TM\oplus \End X$ locally defined by
$(\mu^r,\alpha^A{}_B)\rightarrow (\mu^r,\bar\alpha^A{}_B)$, where
\begin{equation}\label{baralapha}
\bar\alpha^A{}_B=A_r{}^A{}_B(m)\mu^r+\alpha^A{}_B.
\end{equation}

\vfill\eject

\section{\normalsize \textcolor{blue}{Action Lie algebroid Poisson cohomology}}
\label{app:diff}

In this appendix, we present a supergeometric description of the 
action Lie algebroid Poisson double complex $(A_J{}^{*,*}(L),d_{J,L},d_P)$ 
introduced in sect. \ref{sec:LAH*}. The following construction is based on the graded vector bundle
$J^*L[1]\oplus T^*[1]X$ with base $X$. $A_J{}^{*,*}(L)$ 
and $d_{J,L}$ , $d_P$ are then realized as a subspace of functions  on $J^*L[1]\oplus T^*[1]X$ 
and as degree 1 vector fields on $J^*L[1]\oplus T^*[1]X$, respectively.

Denote by $\xi^a$ and $(\gamma^i,\eta_a)$ the base and odd fiber coordinates of 
$J^*L[1]\oplus T^*[1]X$, respectively. Then, a generic function
$\Phi\in C^\infty(J^*L[1]\oplus T^*[1]X)$ has the form
\begin{equation}
\Phi=\sum_{p,q\geq 0}\frac{1}{p!q!}\phi^{(p,q)}{}_{i_1\ldots i_p}{}^{a_1\ldots a_q}(\xi)
\gamma^{i_1}\ldots\gamma^{i_p}\eta_{a_1}\ldots\eta_{a_q}
\end{equation}
with $\phi^{(p,q)}\in \Gamma(\wedge^p(J^*L)^*\otimes\wedge^qTX)$.
Now, define the degree $-1$ vector fields on $J^*L[1]\oplus T^*[1]X$ 
\hphantom{xxxxxxxxxxxxxxxxxxxxxxx}
\begin{equation}
K^r=\partial_a J^r(\xi)\partial_\eta{}^a,
\end{equation}
where $\partial_\eta{}^a=\partial/\partial \eta_a$. So, recalling that
$A_J{}^{p,r}(L)=\Gamma(\wedge^p(J^*L)^*\otimes\wedge^r T^JX)$, it appears that 
$A_J{}^{*,*}(L)$ is identified with the intersection of the kernels of
the $K^r$.

By inspection, one can check that the differentials $d_{J,L}$ , $d_P$ of $A_J{}^{*,*}(L)$ 
are then identified with the degree $1$ vector fields on $J^*L[1]\oplus T^*[1]X$ 
\begin{align}
&d_{J,L}=u_i{}^a(\xi)\gamma^i\partial_a-\partial_bu_i{}^a(\xi)\gamma^i\eta_a\partial_\eta{}^b
-\frac{1}{2}f^k{}_{ij}(J(\xi))\gamma^i\gamma^j\partial_{\gamma k},
\vphantom{\bigg[}
\label{}
\\
&d_P=-P^{ab}(\xi)\eta_a\partial_b+\frac{1}{2}\partial_cP^{ab}(\xi)\eta_a\eta_b\partial_\eta{}^c,
\vphantom{\bigg[}
\label{}
\end{align}
where $\partial_{\gamma i}=\partial/\partial\gamma^i$.
These satisfy the graded commutation relations
\begin{subequations}
\begin{align}
&[d_{J,L},K^r]=\partial_s\rho_i{}^r(J(\xi))\gamma^i K^s,
\vphantom{\bigg[}
\label{}
\\
&[d_P,K^r]=0,
\vphantom{\bigg[}
\label{}
\\
&[d_{J,L},d_{J,L}]=-\partial_rf^k{}_{ij}(J(\xi))u_k{}^a(\xi)\gamma^i\gamma^j\eta_aK^r,
\vphantom{\bigg[}
\label{}
\end{align}
\begin{align}
&[d_{J,L},d_P]=0, \hskip 4.8cm 
\vphantom{\bigg[}
\label{}
\\
&[d_P,d_P]=0,
\vphantom{\bigg[}
\label{}
\end{align}
\end{subequations}
as can be checked using  \eqref{LAA8}, \eqref{LAA16}, \eqref{LAA17}, \eqref{LAA21}, 
\eqref{LAA23}, \eqref{LAA25}. Therefore, $d_{J,L}$ , $d_P$ preserve
the subspace of functions $A_J{}^{*,*}(L)$ and are nilpotent and anticommute on it,
as they should.

\vfill\eject


\end{document}